\documentclass[gmd, manuscript]{copernicus}

\usepackage{float}
\usepackage{amsmath}
\usepackage{listings}
\DeclareCaptionFormat{listing}{#1#2\par} 

\begin{document}
\nolinenumbers  

\title{Consistent Point Data Assimilation in Firedrake and Icepack}


\Author[1,2]{Reuben W.}{Nixon-Hill}
\Author[3]{Daniel}{Shapero}
\Author[1]{Colin J.}{Cotter}
\Author[1]{David A.}{Ham}

\affil[1]{Department of Mathematics, Imperial College London, London, SW7 2AZ}
\affil[2]{Science  and  Solutions  for  a  Changing  Planet  DTP,  Grantham  Institute  for  Climate  Change  and  the  Environment, Imperial College London, London, SW7 2AZ}
\affil[3]{Polar Science Center, Applied Physics Laboratory, University of Washington, 1013 NE}




\correspondence{Reuben W. Nixon-Hill (reuben.nixon-hill10@imperial.ac.uk, \url{https://orcid.org/0000-0001-6226-4640})}

\runningtitle{TEXT}

\runningauthor{TEXT}

\received{}
\pubdiscuss{} 
\revised{}
\accepted{}
\published{}


\firstpage{1}

\maketitle

\begin{abstract}

When estimating quantities and fields that are difficult to measure directly, such as the fluidity of ice, from point data sources, such as satellite altimetry, it is important to solve a numerical inverse problem that is formulated with Bayesian consistency.
Otherwise, the resultant probability density function for the difficult to measure quantity or field will not be appropriately clustered around the truth.
In particular, the inverse problem should be formulated by evaluating the numerical solution at the true point locations for direct comparison with the point data source.
If the data are first fitted to a gridded or meshed field on the computational grid or mesh, and the inverse problem formulated by comparing the numerical solution to the fitted field, the benefits of additional point data values below the grid density will be lost.
We demonstrate, with examples in the fields of groundwater hydrology and glaciology, that a consistent formulation can increase the accuracy of results and aid discourse between modellers and observationalists.

To do this, we bring point data into the finite element method ecosystem as discontinuous fields on meshes of disconnected vertices.
Point evaluation can then be formulated as a finite element interpolation operation (dual-evaluation).
This new abstraction is well-suited to automation, including automatic differentiation.
We demonstrate this through implementation in Firedrake, which generates highly optimised code for solving Partial Differential Equations (PDEs) with the finite element method.
Our solution integrates with dolfin-adjoint/pyadjoint, allowing PDE-constrained optimisation problems, such as data assimilation, to be solved through forward and adjoint mode automatic differentiation.

\end{abstract}


\introduction  
Many disciplines in the earth sciences face a common problem in the lack of observability of important fields and quantities.
In groundwater hydrology, the conductivity of an aquifer is not directly measurable at large scales; in seismology, the density of the earth; and in glaciology, the fluidity of ice.
Nevertheless, these are necessary input variables to the mathematical models that are used to make predictions.
\emph{Inverse problems} or \emph{data assimilation} have us estimating these immeasurables through some mathematical model that relates them to something measurable.
For example, the large scale density (an immeasurable) and displacement (a measurable) of the earth are related through the seismic wave equation.
Combining measurements of displacement or wave travel time from active or passive seismic sources can then give clues to the density structure of the earth.
Computing the most probable estimate of an unobservable results in an optimisation problem with, typically, a Partial Differential Equation (PDE) as a constraint.
This problem is often solved by minimising some model-data misfit metric\footnote{In weather and climate models, this is referred to as \emph{variational} data assimilation.} which includes a regularisation term to keep the problem well formed.

Geoscientists use a wide variety of measurement techniques to study the Earth system. Each of these techniques yields data of different density in space and time.
The number of independent measurements and their accuracy dictates how much information can be obtained about unobservable fields through data assimilation.
The tools used to solve these PDE-constrained optimisation problems do not always allow the varied density in space and time of such measurements to be taken into account.
In such scenarios it can be tempting to create a model-data misfit which treats such measurements as a continuous field: exactly how to create such a field from discrete measurements is a choice the modeller must make.
Alternatively, if a tool allows it, one can create a model-data misfit that is evaluated at the discrete points.
This distinction has important consequences.
For example, if observations are sparse then using a misfit which treats measurements as a continuous field requires significant assumptions.

To use model-data misfits which are evaluated at discrete points we need to be able to perform point evaluation on the fields which we compare to our discrete measurements.
These point evaluations need to integrate with whatever method we are using to solve our optimisation problem: in many cases this requires finding a first or second derivative of the point evaluation operation alongside operations such as solving a PDE.

The models we are interested in are typically numerical discretisations of systems of differential equations to which we find some numerical solution.
Galerkin methods, here referred to generally as finite element methods, are a popular approach with useful properties.
We show how we can integrate point data into the finite element method paradigm of fields on meshes and demonstrate that this can be used to automate point evaluation such that we can automatically solve these minimisation problems.

We demonstrate our approach by implementing it in the finite element library Firedrake \citep{rathgeber_firedrake_2016}.
Firedrake takes symbolic mathematical expressions of PDEs written in the Unified Form Language (UFL) \citep{alnaes_ufl_2012} domain specific language, and generates parallelisable, scalable \citep{betteridge_code_2021} and efficient finite element C code.
It supports a wide array of elements and has been used to build the ocean model Thetis \citep{karna_thetis_2018}, atmospheric dynamical core Gusto \citep{ham_automating_2017}, and glacier flow modelling toolkit Icepack \citep{shapero2021icepack}.
Firedrake integrates with the discrete adjoint generation system dolfin-adjoint/pyadjoint \citep{mitusch_dolfin-adjoint_2019} which we use to compare the two model-data misfit approaches.
We go on to demonstrate its use for influencing experiment design in groundwater hydrology, where measurements are generally very sparse.
Lastly we perform a cross-validation data assimilation experiment in glaciology using Icepack \citep{shapero2021icepack}.
This experiment requires model-data misfit terms which use point evaluations and allows us to infer information about the statistics of our assimilated data.

The paper is laid out as follows.
Sections \ref{sec:finite_element_fields}, \ref{sec:point_data} and \ref{sec:interpolation} describe how we integrate point data with finite element methods while Sect. \ref{sec:implementation} shows our specific Firedrake implementation.
In Sect. \ref{sec:data_assimilation} we return to the topic of data assimilation and pose the question of model-data misfit choice in more detail; we investigate this in Sect. \ref{sec:unknown_conductivity} using our new Firedrake implementation.
The further demonstrations in groundwater hydrology and glaciology can be found in Sections \ref{sec:hydrology} and \ref{sec:ice_shelves}, respectively.

\section{Finite element fields} \label{sec:finite_element_fields}

In finite element methods the domain of interest is approximated by a set of discrete cells known as a mesh $\Omega$.
The solution of a PDE $u$ is then approximated as the sum of a discrete number of functions on the mesh.
Each function is conveniently defined to be a \emph{basis} or \emph{shape} function $\phi$, multiplied by some weight coefficient $w$.
For $N$ weight coefficients and basis functions our approximate solution
\begin{equation}
    u(x) = \sum_{i=0}^{N-1} w_i \phi_i(x) \label{eq:finite_element_fields:def}
\end{equation}
is called a \emph{finite element field}.
The set of basis functions $\{\phi_i(x)\}$ are kept the same for a given mesh but the weights $\{w_i\}$ are allowed to vary to form our given $u(x)$: these weights are referred to as Degrees of Freedom (DoFs).

The set of all possible weight coefficients applied to the basis functions on our mesh is called a finite element Function Space $\text{FS}(\Omega)$, here referred to as a \emph{finite element space}.
Our finite element field is therefore a member of our finite element space 
\begin{equation}
    u \in \text{FS}(\Omega).
\end{equation}

\begin{figure}[hbtp]
    \centering
    \includegraphics[width=\linewidth]{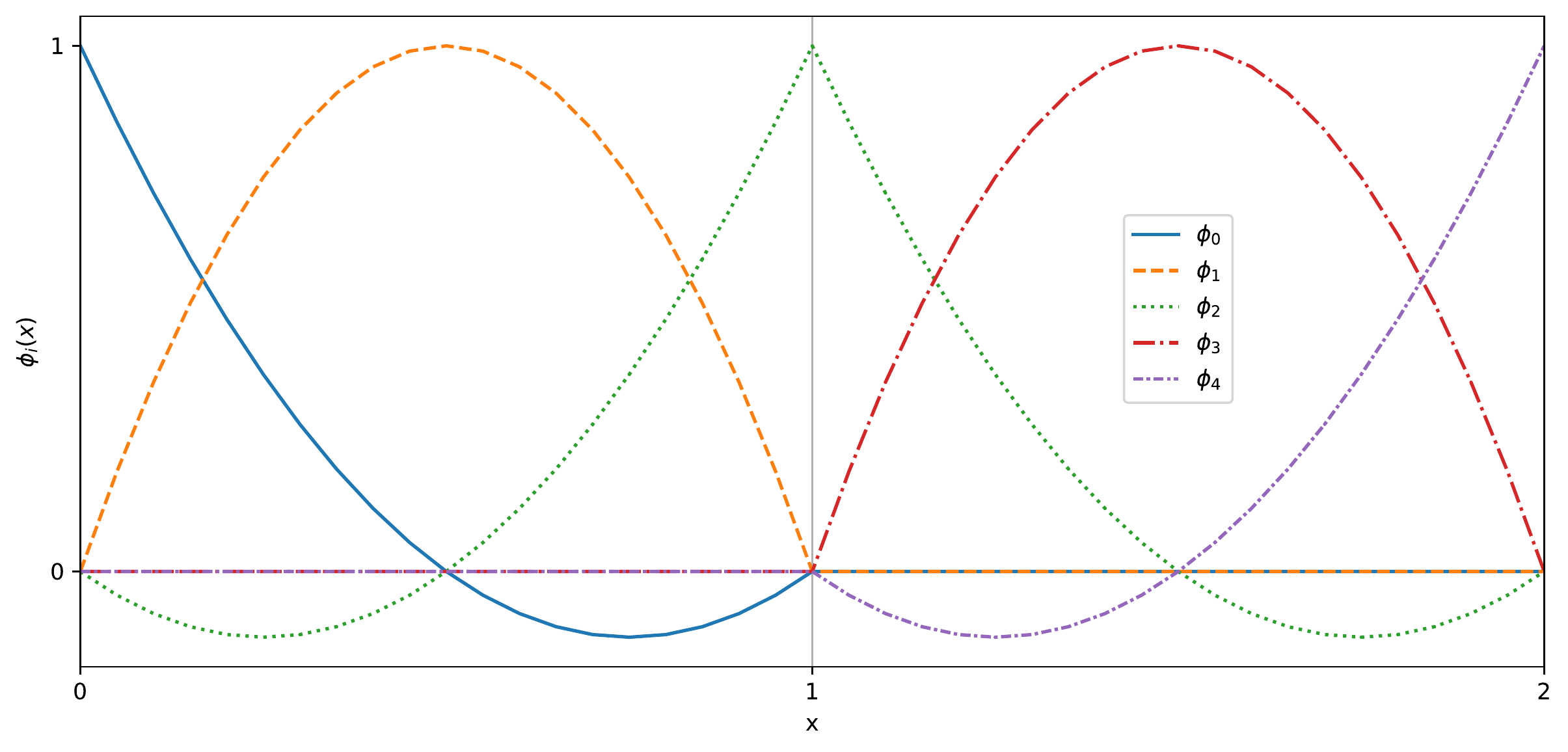}
    \caption{
        Second order Lagrange basis polynomials on an interval $[0, 2]$ meshed into two cells, $[0, 1]$ and $[1, 2]$.
        Only $\phi_2$ is nonzero in both cells.
        }
    \label{fig:diagrams:2nd_order_lagrange_line_mesh}
\end{figure}

Finite element spaces are grouped by their definitions on particular mesh cell shapes with precisely defined basis functions in each case.
A popular choice are piecewise polynomials, such as the second order continuous Lagrange polynomials
\begin{equation}
    u \in \text{P2CG}(\Omega)
\end{equation}
where P2CG stands for Polynomial degree 2 Continuous Galerkin and `continuous' refers to continuity of the field between mesh cells.
For a simple line domain $[0, 2]$ discretised into 2 cells $[0, 1]$ and $[1, 2]$ we have a total of $N=5$ basis functions and weight coefficients (DoFs).
The basis functions are shown in Fig. \ref{fig:diagrams:2nd_order_lagrange_line_mesh}.
Our finite element field $u$ is given by Eq. \ref{eq:finite_element_fields:def}: $u(x) = x^2$ has weights $w_0 = 0$, $w_1 = 0.25$, $w_2 = 1$, $w_3 = 2.25$ and $w_4 = 4$.\footnote{All polynomials of 2nd order below can be represented exactly in this finite element space.}

The consistency of the functions across each cell is clear in this example.
We have 3 weight coefficients and 3 basis functions which are nonzero in each cell in a repeating pattern.
Most finite element libraries therefore calculate weight coefficients cell-by-cell, with the weight of any shared nonzero basis functions (here only $\phi_2$) either calculated twice or passed to neighbouring cells.

There are two key points here.
Firstly, all finite element fields are members of finite element spaces defined on a mesh.
Secondly, our approximated solution (Eq. \ref{eq:finite_element_fields:def}) is a field.
As long as this field is continuous we know its values unambiguously.
We can therefore evaluate the solution at the location of any measurement so long as that location is on the mesh.
We will return to finite element spaces with discontinuities later.

\section{Point data} \label{sec:point_data}

\begin{figure}[hbtp]
    \centering
    \includegraphics[width=0.33\linewidth]{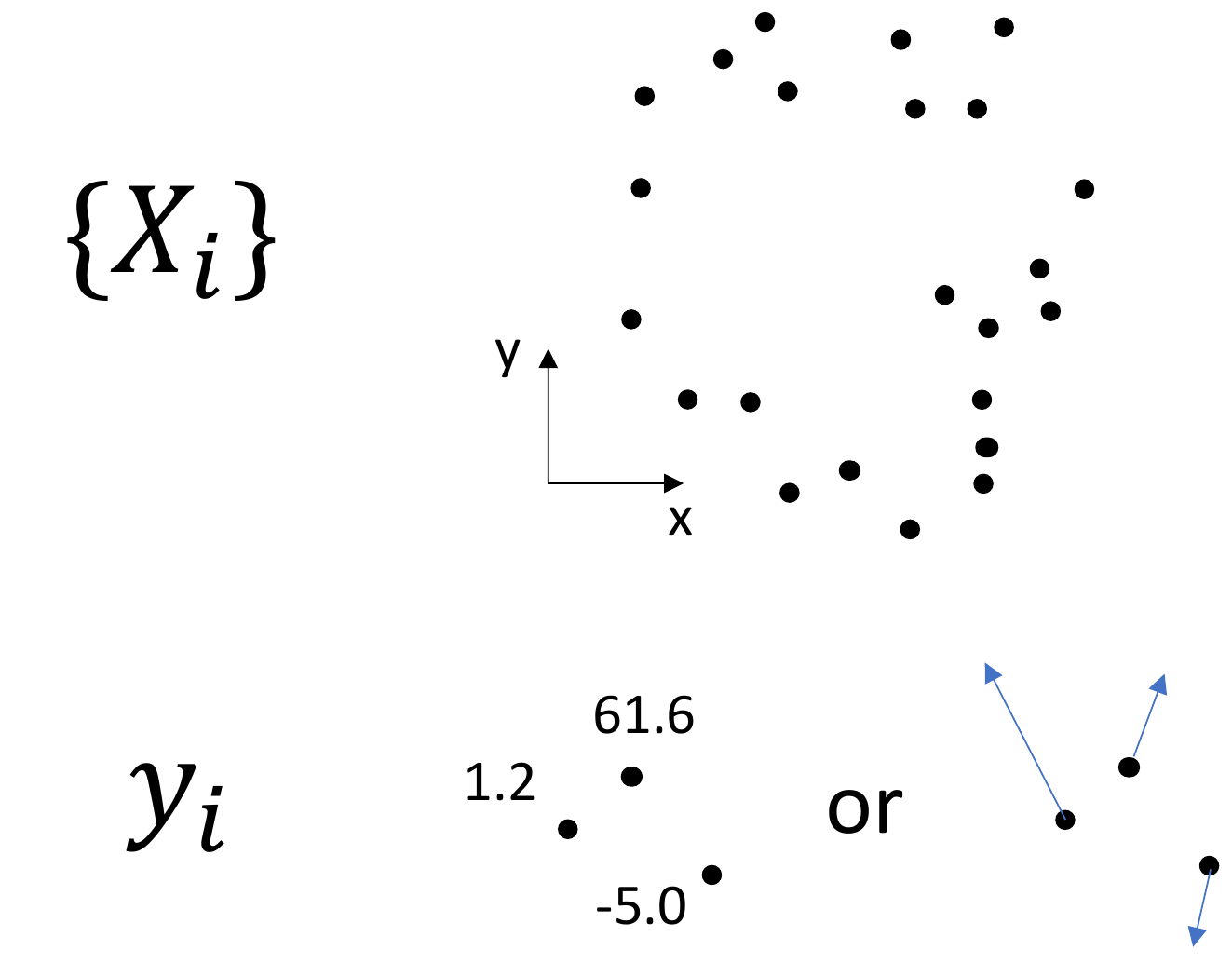}
    \caption{
        Point data consist of (a) a set of $N$ spatial coordinates $\{X_i\}_{i=0}^{N-1}$ (a `point cloud') and (b) the scalar, vector or (not shown) tensor values $\{y_i(X_i)\}_{i=0}^{N-1}$ at those coordinates.
        Maintaining this distinction is key when trying to form a rigorous way of handling point data for any numerical method, such as the finite element method, which separates the idea of domain and values on the domain.}
    \label{fig:diagrams:point_data_definition}
\end{figure}
To integrate point data with our paradigm of finite element fields being members of finite element spaces defined over meshes we need to look carefully at what we mean by point data.
Point data can be separated into two parts: (a) the locations $\{X_i\}$ of the $N$ data points at a given time (a point cloud) and (b) the $N$ values $\{y_i(X_i)\}_{i=0}^{N-1}$ associated with the point cloud (see Fig. \ref{fig:diagrams:point_data_definition}).
Finite element fields have a similar distinction: (a) the discretised shape of the domain of interest (the mesh $\Omega$) and (b) the values associated with that mesh (the weights applied to the basis functions).

Applying the finite element distinction to the locations of the data (a) suggests a `point cloud mesh' formed of $N$ disconnected vertices at each location $X_i$:
\begin{equation}
    \Omega_v = \{X_i\}_{i=0}^{N-1}.
\end{equation}
We refer to this as a `vertex-only mesh'.
This is an unusual mesh: a vertex has no extent (it is topologically zero dimensional) but exists at each location $X_i$ in a space of geometric dimension $\text{dim}(X_i)$.
Fortunately meshes with topological dimension less that their geometric dimension are not unusual: 2D meshes of the surface of a sphere in 3D space are commonly used to represent the surface of the earth.
Such domains are typically called `immersed manifolds'.
Disconnected meshes are also not unheard of: the software responsible merely needs to be able to iterate over all the cells of the mesh.
In this case each cell is a vertex $X_i$.
We can therefore legitimately construct such a mesh.

We now need to consider the values $\{y_i(X_i)\}_{i=0}^{N-1}$ (b).
Only one value, be it scalar, vector or tensor, can be given to each cell (i.e. each point or vertex).
Fortunately a finite element space for this case exists: the space of zero order discontinuous Lagrange polynomials
\begin{equation}
    y \in \text{P0DG}(\Omega_v)
\end{equation}
where P0DG stands for Polynomial degree 0 Discontinuous Galerkin.
Here $y$ is a single discontinuous field which contains all of our point data values at all of our point locations (mesh vertices)
\begin{equation}
    y(x) = 
    \begin{cases}
        y(X_i) & \text{if} \ x = X_i, \\ 
        \text{undefined} & \text{elsewhere}.
    \end{cases}
\end{equation}
Integrating this over our vertex-only mesh $\Omega_v$ gives
\begin{equation}
    \int_{\Omega_v} y(x) dx = \sum_{i=0}^{N-1} y(X_i) \quad \forall y \in \text{P0DG}(\Omega_v) \label{eq:point_data:int_sum_equiv}
\end{equation}

This definition lets us directly reason about point data in finite element language and yield useful results.
So long as the locations of our vertices $X_i$ of our vertex-only mesh $\Omega_v$ are within the domain of our `parent' mesh
\begin{equation}
    \Omega_v \subset \Omega \label{eq:point_data:mesh_immersion}
\end{equation}
then we can go from some field $u$ in some finite element space defined on our parent mesh
\begin{equation}
    u \in \text{FS}(\Omega) \label{eq:point_data:parent_func_space}
\end{equation}
to one defined on our vertex-only mesh
\begin{equation}
    u_v \in \text{P0DG}(\Omega_v) \label{eq:point_data:vom_func_space}
\end{equation}
by performing point evaluations at each vertex location $u(X_i) \ \forall \ i$.

The operator for this can be formulated as finite element interpolation (known as `dual functional evaluation' or simply `dual evaluation') into $\text{P0DG}$, i.e.
\begin{equation}
     \mathcal{I}_{\text{P0DG}(\Omega_v)}(; u) : \text{FS}(\Omega) \to \text{P0DG}(\Omega_v) \label{eq:point_data:interpolation_op}
\end{equation}
such that
\begin{equation}
     \mathcal{I}_{\text{P0DG}(\Omega_v)}(; u) = u_v.
\end{equation}
This operator is linear in $u$ which we denote by a semicolon before the argument.
The construction of this operator is described in the next section.

\section{Point evaluation as an interpolation operator} \label{sec:interpolation}

In most implementations of finite element methods we have a set of `global' coordinates covering our meshed domain and a set of `local' coordinates defined on some reference cell.
For each mesh cell we transform from global to local coordinates, perform an operation, then transform our result back.

This is a rough outline of the definition of interpolation found in Sect. 3.3 of \cite{brenner_mathematical_2008} using our notation.
For a reference cell $\mathcal{K}$, local interpolation $\mathcal{I_{P(K)}}$ for a set of $k$ local basis functions $\mathcal{P} = \text{span}(\{\Tilde{\phi}_i\}_0^{k-1})$ of some locally defined field $\Tilde{f}$ is given by the linear operator
\begin{equation}
    \big[\mathcal{I_{P(K)}}(; \Tilde{f})\big](\hat{x}) = \sum_{j=0}^{k-1} \Tilde{\phi'_j}(; \Tilde{f}) \Tilde{\phi_j}(\hat{x}) \label{eq:interpolation:general_definition}
\end{equation}
where $\{\Tilde{\phi'_i}\}_0^{k-1}$ are the local dual basis linear functionals (the span of which are the nodes $\mathcal{N}$) and $\Tilde{x}$ are our local coordinates.
This uses Ciarlet's triple formulation $(\mathcal{K}, \mathcal{P}, \mathcal{N})$ \citep{ciarlet_finite_2002} as the definition of a finite element.
Global interpolation over the entire mesh $\Omega$ for the complete finite element space $\text{FS}(\Omega)$, which we denote $\mathcal{I}_{\text{FS}(\Omega)}$, is the local application (i.e. transformed to local reference coordinates) of $\mathcal{I_{P(K)}}$ to the globally defined field $f \in \text{FS}(\Omega)$.

Dual basis functionals are strictly a mapping from members of $\mathcal{P}$ to a scalar: i.e. the global interpolation operator for a given finite element space $\text{FS}(\Omega)$ is defined in all cases for fields in that space ($\mathcal{I}_{\text{FS}(\Omega)} : \text{FS}(\Omega) \to \text{FS}(\Omega)$).
It is not unusual, however, for finite element libraries to allow interpolation from fields which are defined in another finite element space so long as the geometric dimension and meshed domains are consistent\footnote{One usually also needs the finite element spaces to be continuous at cell boundaries for the operator to be well defined everywhere but this is not always explicitly checked for.}.
Both Firedrake and FEniCS \citep{logg_efficient_2009, alnaes_fenics_2015} allow this.
It works because most dual basis functionals involve either evaluating a point on the cell $\mathcal{K}$, doing some integration over that cell, or evaluating a derivative component on it.
So long as this can be done with sufficient accuracy in the finite element space we are interpolating from, we can perform our interpolation and get results which have properties which we expect the new space to give us, such as having particular values at particular points (for more see \cite{maddison_directional_2012}).

For $u_v = \mathcal{I}_{\text{P0DG}(\Omega_v)}(; u)$ where $u(x) \in \text{FS}(\Omega)$ we require, at each vertex cell $X_i$ of our vertex-only mesh $\Omega_v$, the point evaluation $u(X_i)$.
This implies that, for each $X_i$, we require a single local dual basis functional (i.e. $k=1$ in Eq. \ref{eq:interpolation:general_definition}) which performs the necessary point evaluation.

Following the pattern of Eq. \ref{eq:interpolation:general_definition}, we ought to define a fixed reference vertex $\Tilde{X}$ and transform $u$ to some $\Tilde{u}$ such that 
\begin{equation}
    \Tilde{\phi'}(\Tilde{u}) = \Tilde{u}(\Tilde{X}).
\end{equation}
then transform back to the global coordinates of our vertex-only mesh $\Omega_v$ such that, in each case, we perform $u(X_i)$.
This would give us a local basis function $\Tilde{\phi}(\Tilde{x}) = 1$ when $\Tilde{x} = \Tilde{X}_i$.

We can equivalently let the reference vertex, and therefore our functional, vary for each $X_i$
\begin{equation}
    \Tilde{\psi'_i}(\hat{u}) = \hat{u}(\hat{X}_i)
\end{equation}
where $\hat{u}$ is our locally defined field on the reference cell of our \emph{parent} mesh $\Omega$.
Now for each vertex $X_i$ in $\Omega_v$, $\hat{X}_i$ is its equivalent location in the reference cell of the parent mesh $\Omega$.
This gives us a local basis function $\Tilde{\psi_i}(\hat{x}) = 1$ when $\hat{x} = \hat{X}_i$ which is equivalent to $\Tilde{\phi}(\Tilde{x})$ since, after transforming back to global coordinates, they both equal 1 at $X_i$.
The global interpolation operator then has us doing the following for each vertex cell $X_i$ in our vertex-only mesh $\Omega_v$:
\begin{enumerate}
    \item finding the cell of the parent mesh $\Omega$ that $X_i$ resides in, \label{itm:interpolation:1}
    \item finding the equivalent reference coordinate $\hat{X}_i$ in that cell, \label{itm:interpolation:2}
    \item transforming our field $u$ to reference coordinates giving $\hat{u}$, \label{itm:interpolation:3}
    \item performing the point evaluation\footnote{For vector or tensor valued function spaces, this becomes the inner product of the point evaluation with a Cartesian basis vector or tensor respectively.} $\hat{u}(\hat{X}_i)$ and \label{itm:interpolation:4}
    \item transforming the result back to global coordinates giving $u(X_i)$.\label{itm:interpolation:5}
\end{enumerate}

This operation formalises the process of point evaluation with everything remaining a finite element field defined on a mesh.
These fields can have concrete values or be symbolic unknowns.
If the symbolic unknown is a point, we can now express that in the language of finite elements.
Whilst it is not the topic of this paper, we can now, for example, express point forcing expressions as
\begin{equation}
    \int_{\Omega_v} \mathcal{I}_{\text{P0DG}(\Omega_v)}(; f(x)) dx = \sum_{i=0}^{N-1} f(x_i) = \sum_{i=0}^{N-1} \int_\Omega f(x) \delta(x - x_i) dx. \label{eq:interpolation:delta_equiv}
\end{equation}

Given later discussion, note here that what we call an `interpolation' operation is exact and, excepting finite element spaces with discontinuities, unique: we get the value of $u$ at the points $\{X_i\}$ on the new mesh $\Omega_v$.
We are able to perform this exact interpolation because $u$ is a field which has a value everywhere.
A concrete example is considered in Sect. \ref{sec:demonstrations}.

\section{Firedrake implementation} \label{sec:implementation}

\begin{lstlisting}[language=Python, label={lst:implementation:firedrake_example}, caption={An example point evaluation in Firedrake: the last three lines are our new functionality. In Firedrake, as in much finite element literature, what we call fields and finite element spaces are known as functions and function spaces respectively.\captionnewline}]
from firedrake import *

def poisson_point_eval(coords):
    """Solve Poisson's equation on a unit square for a random forcing term
    with Firedrake and evaluate at a user-specified set of point coordinates.

    Parameters
    ----------
    coords: numpy.ndarray
        A point coordinates array of shape (N, 2) to evaluate the solution at.

    Returns
    -------
    firedrake.function.Function
        A field containing the point evaluatations.
    """
    omega = UnitSquareMesh(20, 20)
    P2CG = FunctionSpace(omega, family="CG", degree=2)
    u = Function(P2CG)
    v = TestFunction(P2CG)

    # Random forcing Function with values in [1, 2].
    f = RandomGenerator(PCG64(seed=0)).beta(P2CG, 1.0, 2.0)

    F = (inner(grad(u), grad(v)) - f * v) * dx
    bc = DirichletBC(P2CG, 0, "on_boundary")
    solve(F == 0, u, bc)

    omega_v = VertexOnlyMesh(omega, coords)
    P0DG = FunctionSpace(omega_v, "DG", 0)
    return interpolate(u, P0DG)
\end{lstlisting}

Firedrake code for specifying and solving PDEs barely differs from the equivalent mathematical expressions.
As an example see Listing \ref{lst:implementation:firedrake_example}, in which we solve Poisson's equation $-\nabla^2 u = f$ under strong (Dirichlet) boundary conditions $u = 0$ on the domain boundary, in just 8 lines of code.
We specify our domain $\Omega$ and finite element space (here called a function space) to find a field solution $u \in \text{P2CG}(\Omega)$ (here called a function).
Since we use the finite element method we require a weak formulation of Poisson's equation
\begin{equation}
    \int_\Omega \nabla u \cdot \nabla v - f v \ dx = 0 \ \ \forall v \in \text{P2CG}(\Omega)
\end{equation}
where $v$ is called a test function\footnote{For those unfamiliar with the finite element method, this is standard nomenclature.}.
The specification of our PDE, written in UFL on line 25, is identical to the weak formulation.
We apply our boundary conditions and then solve.
Firedrake uses PETSc \citep{petsc-web-page, petsc-efficient} to solve both linear and nonlinear PDEs\footnote{Our solver functions are more flexible than this simple example suggests. 
For example, PETSc solver options can be passed from Firedrake to PETSc.
See the PETSc manual \citep{petsc-user-ref} and Firedrake project website \url{https://www.firedrakeproject.org/} for more information}.

Having found a solution we then set about point evaluating it.
The final three lines of Listing \ref{lst:implementation:firedrake_example} follow the mathematics of Sect. \ref{sec:point_data} and Sect. \ref{sec:interpolation}.
Our vertex-only mesh $\Omega_v$ is immersed in our parent mesh $\Omega$ (Eq. \ref{eq:point_data:mesh_immersion}) using a new \verb|VertexOnlyMesh| constructor. 
The finite element space $\text{P0DG}(\Omega_v)$ is then created (Eq. \ref{eq:point_data:vom_func_space}) using existing Firedrake syntax.
Lastly the cross mesh interpolation operation (Eq. \ref{eq:point_data:interpolation_op}) is performed; this again uses existing Firedrake syntax.

The vertex-only mesh implementation uses the DMSWARM point-cloud data structure in PETSc. We build our $\text{P0DG}$ finite element space on top of \verb|VertexOnlyMesh| in the same way we would any other mesh.\footnote{Other mesh types in Firedrake use the DMPlex data structure rather than DMSWARM as explained in \cite{lange_efficient_2016}.
The construction of a finite element space (a function space), proceeds exactly as shown in Fig. 2 of Lange et al. but with DMPlex now a DMSWARM.}
Our interpolation operation follows the steps laid out in Sect. \ref{sec:interpolation}.

Our implementation works seamlessly with Firedrake's MPI parallelism \citep{mpi40}.
Firedrake performs mesh domain decomposition when run in parallel: vertex-only mesh points decompose across ranks as necessary.
Where points exist on mesh partition boundaries a voting algorithm ensures that only one MPI rank is assigned the point.
The implementation supports both very sparse and very dense points as we will demonstrate in Sect. \ref{sec:demonstrations}.

Where solutions on parent meshes have discontinuities we allow point evaluation wherever it is well defined.
Solutions on parent meshes from discontinuous Galerkin finite element spaces, which are not well defined on the cell boundaries, can still be point evaluated here: the implementation picks which cell a boundary point resides in.

All interpolation operations in Firedrake can be differentiated using a Firedrake extension to the dolfin-adjoint/pyadjoint package which allows us to solve PDE-constrained optimisation problems.
This was a key motivating factor for turning point evaluation into an interpolation operation as we will go on to show.

\section{Assimilating point data} \label{sec:data_assimilation}

We start with some model
\begin{equation}
    F(u, m) = 0
\end{equation}
where $m$ is a set of parameters; $u$ is our solution; and $F$ the equation, such as a PDE, that relates $m$ and $u$.
The point data to assimilate is
\begin{equation}
    u_{\text{obs}}^i \ \text{at} \ X_i.
\end{equation}
Finding the parameters $m$ which give us these data is called solving an inverse problem.

A typical formulation involves running the model for some $m$ and employing a model-data misfit metric to see how closely the output $u$ matches the data $\{u_{\text{obs}}^i\}$.
We then minimise the model-data misfit metric: the parameters $m$ are updated, using a method such as gradient descent, and the model is run again.
This is repeated until some optimum has been reached.

The model-data misfit metric is a functional (often called the `objective function') $J$ with a typical form being
\begin{equation}
    J = J_{\text{model-data misfit}} + J_{\text{regularisation}}.
\end{equation}
The regularisation can be thought of as a prior guess at the properties our solution should have.
Often this uses known properties of the physics of the model (such as some smoothness requirement) and, in general, ensures that the problem is well posed given limited, typically noisy, measurements of the true field $u$.

A note on nomenclature: $J_{\text{model-data misfit}}$ is, in certain circumstances, known as a `loss function' and the term `cost function' can be used to refer to $J_{\text{model-data misfit}}$ or the whole of $J$.
These terms are particularly prevalent in machine learning literature.
Here we will stick to referring to all as `functionals'.

A key question to ask here is ``what metric should we use for the model-data misfit?''
One approach, taken for example by \citet{shapero2016basal}, is to perform a field reconstruction: we extrapolate from our set of observations to get an approximation of the continuous field we aimed to measure.
This reconstructed field $u_{\text{interpolated}}$ is then compared with the solution field $u$
\begin{equation}
    J_{\text{model-data misfit}}^{\text{field}} = \Vert u_{\text{interpolated}} - u \Vert_N
\end{equation}
where $\Vert \cdot \Vert_N$ is some norm.
We call the extrapolated reconstruction $u_{\text{interpolated}}$ since, typically, this relies on some `interpolation' regime found in a library such as SciPy \citep{2020SciPy-NMeth} to find the values between measurements. 
As we will see when we return to these methods in Sect. \ref{sec:unknown_conductivity}, $J_{\text{model-data misfit}}^{\text{field}}$ is not unique since there is no unique $u_{\text{interpolated}}$ field.
The method used to create $u_{\text{interpolated}}$ is up to the modeller and is not always reported.

An alternative metric is to compare the point evaluations of the solution field $u(X_i)$ with the data $u_{\text{obs}}^i$
\begin{equation} \label{eq:data_assimilation:before_change}
    J_{\text{model-data misfit}}^{\text{point}} = \Vert u_{\text{obs}}^i - u(X_i)  \Vert_N \ \forall \ i.
\end{equation}
Importantly, $J_{\text{model-data misfit}}^{\text{point}}$ is, with the previously noted discontinous Galerkin exception, unique and independent of any assumptions made by the modeller.

It is the difference between minimising $J_{\text{model-data misfit}}^{\text{field}}$ and $J_{\text{model-data misfit}}^{\text{point}}$ which we investigate here.
Previously we could generate code to minimise a functional containing $J_{\text{model-data misfit}}^{\text{field}}$ using Firedrake and dolfin-adjoint/pyadjoint.
Dolfin-adjoint/pyadjoint performs tangent-linear and adjoint mode Automatic Differentiation (AD) on Firedrake operations\footnote{The current implementation of dolfin-adjoint/pyadjoint is a general AD tool for the python language using algorithms from \cite{naumann_art_2011}.
This is described in detail in \cite{mitusch_algorithmic_2018}.
Firedrake includes a wrapper around dolfin-adjoint/pyadjoint which allows AD of Firedrake operations such as interpolation.}, including finding the solutions to PDEs\footnote{This requires automated formulation of adjoint PDE systems.
See \cite{farrell_automated_2013} and Sect. 1.4 of \cite{schwedes_mesh_2017} for more detail.} and performing interpolation.
Point evaluation operations have been a notable exception.
Now that we can automatically differentiate point evaluation operations by recasting them as interpolations we can investigate minimising a functional which contains $J_{\text{model-data misfit}}^{\text{point}}$.
As Listing \ref{lst:unknown_conductivity:minimization_code} in Sect. \ref{sec:unknown_conductivity} shows, this requires just a few lines of code.
To the author's knowledge, the technology needed to minimise $J_{\text{model-data misfit}}^{\text{point}}$ using automated code generating finite element method technology has, until now, not been readily possible.

\section{Demonstrations} \label{sec:demonstrations}

\subsection{Unknown conductivity} \label{sec:unknown_conductivity}

We start with the $L^2$ norm for our two model-data misfit functionals
\begin{equation}
    J_{\text{model-data misfit}}^{\text{field}} = \int_{\Omega} ( u_{\text{interpolated}} - u )^2 dx
\end{equation}
and
\begin{equation} \label{eq:conductivity:Jmisfitpoint}
    J_{\text{model-data misfit}}^{\text{point}} = \int_{\Omega_v} ( u_{\text{obs}} - \mathcal{I}_{\text{P0DG}(\Omega_v)}(u) )^2 dx
\end{equation}
where $u_\text{obs} \in \text{P0DG}(\Omega_v)$.
Since integrations are equivalent to the sums of point evaluations in $\text{P0DG}(\Omega_v)$ (Eq. \ref{eq:point_data:int_sum_equiv}), the $L^2$ norm is the same as the euclidean ($l^2$) norm.
Our misfit is therefore evaluated as
\begin{equation} \label{eq:conductivity:Jmisfitpointsum}
    J_{\text{model-data misfit}}^{\text{point}} = \sum_{i=0}^{N-1} ( u_{\text{obs}}^i - u(X_i) )^2.
\end{equation}

We will apply this to the simple model
\begin{equation}
    -\nabla\cdot k\nabla u = f
\end{equation}
for some solution field $u$ and known forcing term $f = 1$ with conductivity field $k$ under strong (Dirichlet) boundary conditions
\begin{equation}
    u = 0 \ \text{on} \ \Gamma
\end{equation}
where $\Gamma$ is the domain boundary.
We assert conductivity $k$ is positive by
\begin{equation}
  k = k_0e^q
\end{equation}
with $k_0 = 0.5$.

Our inverse problem is then to infer, for our known forcing field term $f$, the log-conductivity field $q$ using noisy sparse point measurements of our solution field $u$.
This example has the advantage of being relatively simple whilst having a control term $q$ which is nonlinear in the model. 

To avoid considering model discretisation error we generate the log-conductivity field $q_{\text{true}}$ in the space of order 2 continuous Lagrange polynomials ($\text{P2CG}$) in 2D on a $32 \times 32$ unit-square mesh $\Omega$ with 2048 triangular cells.
We then solve the model on the same mesh to get the solution field $u_{\text{true}} \in \text{P2CG}(\Omega)$. 
$N$ point measurements $\{u_{\text{obs}}^i\}_0^{N-1}$ at coordinates $\{X_i\}_0^{N-1}$ are sampled from $u_{\text{true}}$ and Gaussian random noise with standard deviation $\{\sigma_i\}_0^{N-1}$ is added to each measurement.

We use a smoothing regularisation on our $q$ field which is weighted with a parameter $\alpha$.
This helps to avoid over-fitting to the errors in $u_{\text{obs}}$ which are introduced by the Gaussian random noise.
We now have two functionals which we minimise
\begin{equation} \label{eq:conductivity:J}
    J[u, q] = \underbrace{
                          \int_{\Omega_v} ( u_{\text{obs}} - \mathcal{I}_{\text{P0DG}(\Omega_v)}(u) )^2 dx
                          }_{J_{\text{model-data misfit}}^{\text{point}}} + 
              \underbrace{
                          \alpha^2\int_\Omega|\nabla q|^2 dx
                          }_{J_{\text{regularisation}}}
\end{equation}
and
\begin{equation} \label{eq:conductivity:Jprime}
    J'[u, q] = \underbrace{
                          \int_{\Omega} ( u_{\text{interpolated}} - u )^2 dx
                          }_{J_{\text{model-data misfit}}^{\text{field}}} + 
              \underbrace{
                          \alpha^2\int_\Omega|\nabla q|^2 dx
                          }_{J_{\text{regularisation}}}.
\end{equation}

Each available method in SciPy's interpolation library are tested to find $u_{\text{interpolated}}$:
\begin{itemize}
    \item $u_\text{interpolated}^\text{near.}$ using \verb|scipy.interpolate.NearestNDInterpolator|,
    \item $u_\text{interpolated}^\text{lin.}$ using \verb|scipy.interpolate.LinearNDInterpolator|,
    \item $u_\text{interpolated}^\text{c.t.}$ using \verb|scipy.interpolate.CloughTocher2DInterpolator| with \verb|fill_value = 0.0| and
    \item $u_\text{interpolated}^\text{gau.}$ using \verb|scipy.interpolate.Rbf| with Gaussian radial basis function.
\end{itemize}
Note that since $u_{\text{interpolated}} \in \text{P2CG}(\Omega)$ each of 6 degrees of freedom per mesh cell has to have a value estimated given the available $u_{\text{obs}}$. 

The estimated log-conductivity $q_{\text{est}}$ which minimise the functionals are found using gradient decent by generating code for the adjoint of our model using dolfin-adjoint/pyadjoint then using the Newton-CG minimiser from the \verb|scipy.optimize| library.
To use Newton-CG the ability to calculate Hessian-vector products for Firedrake interpolation operations was added to Firedrake's pyadjoint plugin modules.\footnote{Pyadjoint uses a forward-over-reverse scheme to calculate Hessian-vector products via an implementation of Eq. 3.8 in \citet{naumann_art_2011}.}

To try and balance the relative weights of the model-data misfit and regularisation terms in $J$ and $J'$ we perform an l-curve analysis \citep{hansenUseLCurveRegularization1993} to find a suitable $\alpha$ following the example of \citet{shapero2016basal}.
The l-curves were gathered for $N = 256$ randomly chosen point measurements with the resultant plots shown in Fig. \ref{fig:conductivity:l-curves-Jprime} and Fig. \ref{fig:conductivity:l-curves-J}.
For low $\alpha$, $J^{\text{field}}_{\text{misfit}}$ stopped being minimised and solver divergences were seen due to the problem becoming ill formed.
$\alpha = 0.02$ was therefore chosen for each method.
For consistency $\alpha = 0.02$ was also used for $J$.

An extract of the Firedrake and dolfin-adjoint/pyadjoint code needed to minimise $J$ is shown in Listing \ref{lst:unknown_conductivity:minimization_code}.
Once again, our Firedrake expression for $J$ in the code is the same as the mathematics in Eq. \ref{eq:conductivity:J} given that \verb|assemble| performs integration over the necessary mesh.
The last 3 lines are all that are required to minimise our functional with respect to $q$, with all necessary code being generated.
Note that we require a reduced functional (\verb|firedrake_adjoint.ReducedFunctional|) since the optimisation problem depends on both $q$ and $u(q)$: for a thorough explanation see Sect. 1.4 of \cite{schwedes_mesh_2017}.

\begin{lstlisting}[language=Python, label={lst:unknown_conductivity:minimization_code}, caption={Firedrake code for expressing J (Eq. \ref{eq:conductivity:J}) and dolfin-adjoint/pyadjoint code (inside a Firedrake wrapper) for minimizing it with respect to $q$. The omitted PDE solve code is very similar to that in Listing \ref{lst:implementation:firedrake_example}; for more see code and and data availability.\captionnewline}]
from firedrake import *
import firedrake_adjoint  # This is Firedrake's dolfin-adjoint/pyadjoint plugin module

# Import our noisy samples of the true u
u_obs_coords = ...
u_obs_vals = ...

# Solve PDE with a guess for q giving an initial u
...

omega_v = VertexOnlyMesh(omega, u_obs_coords)
P0DG = FunctionSpace(omega_v, 'DG', 0)
u_obs = Function(P0DG)
u_obs.dat.data[:] = u_obs_vals

J_misfit = assemble((u_obs - interpolate(u, P0DG))**2 * dx)
alpha = Constant(0.02)
J_regularisation = assemble(alpha**2 * inner(grad(q), grad(q)) * dx)
J = J_misfit + J_regularisation

q_hat = firedrake_adjoint.Control(q)
J_hat = firedrake_adjoint.ReducedFunctional(J, q_hat)
q_min = firedrake_adjoint.minimize(J_hat, method='Newton-CG')
\end{lstlisting}

\subsubsection{Posterior consistency} \label{sec:demonstrations:posterior}
We expect that the error in our solution, when compared to the true solution, will always decrease as we increase the number of points we are assimilating.
From a Bayesian point of view, his is known as posterior consistency: under appropriate assumptions, in a well-posed Bayesian inverse problem the posterior distribution should concentrate around the true values of the estimated quantity \citep{schwartz1965bayes}.  
The regularisation we choose and the weighting we give it encode information about our assumed prior probability distribution of $q$ before we start assimilating data (adding observations).

Take, for example, the regularisation used in this problem
\begin{equation}
    \alpha^2 \int_\Omega|\nabla q|^2dx.
\end{equation}
This asserts a prior that the solution $q$ which minimises $J$ and $J'$ should be smooth and gives a weighting $\alpha$ to the assertion.
If we have posterior consistency, the contribution of increasing numbers of measurements $u_{\text{obs}}$ should increase the weighting of our data relative to our prior and $q_{\text{est}}$ should converge towards the true solution $q_{\text{true}}$.
It is clear that this ought to happen for $J$, the point evaluation approach, since increasing $N$ increases the number of terms in the model-data misfit sum (i.e. the integral over $\Omega_v$ gets bigger: see the equivalency of Eq. \ref{eq:conductivity:Jmisfitpoint} and Eq. \ref{eq:conductivity:Jmisfitpointsum}).
There is no such mechanism for $J'$, the field reconstruction approaches, since adding more data merely ought to cause $u_{\text{interpolated}}$ to approach $u_{\text{true}}$ without increasing the relative magnitude of the misfit term.

\begin{figure}[hbtp]
    \centering
    \includegraphics[width=0.6\linewidth]{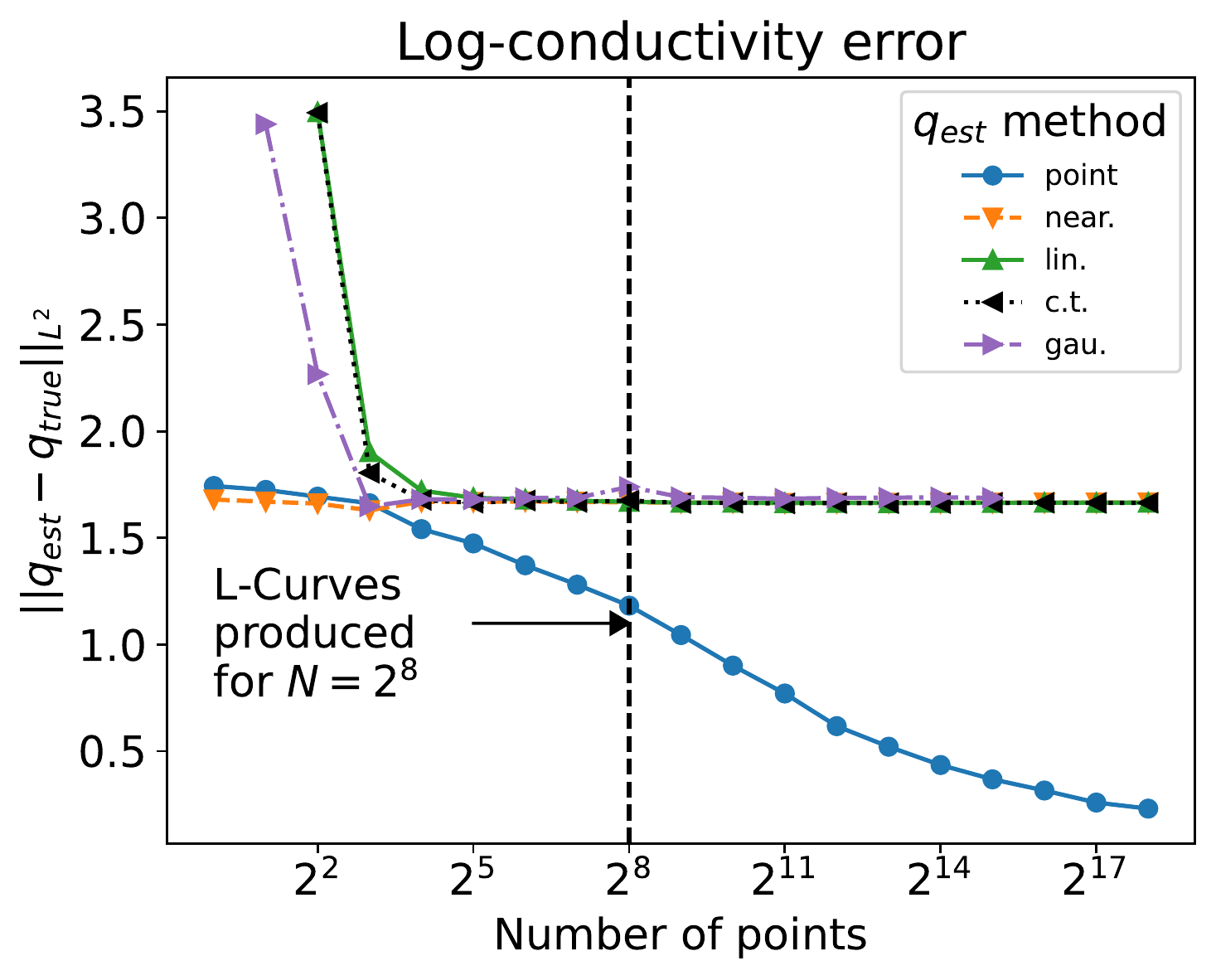}
    \caption{
        Error change as number of points $N$ is increased for minimising $J$ ($u_\text{interpolated}$ and $q_\text{est}$ method `point' - see Eq. \ref{eq:conductivity:J}) and $J'$ (the other lines - see Eq. \ref{eq:conductivity:Jprime}) with different methods for estimating $u_{\text{interpolated}}$ where $\alpha = 0.02$ throughout (see main text for justification).
        The l-curves for $\alpha = 0.02$ with $N = 256$ are shown in Fig. \ref{fig:conductivity:l-curves-Jprime} and Fig. \ref{fig:conductivity:l-curves-J}.
        Not all methods allowed $u_{\text{interpolated}}$ to be reconstructed either due to there being too few point measurements or the interpolator requiring more system memory than was available.}
    \label{fig:conductivity:posterior:l2norms}
\end{figure}

Figure \ref{fig:conductivity:posterior:l2norms} demonstrates that the problem formulated with $J$, the point evaluation approach, both demonstrates posterior consistency and produces a $q_{\text{est}}$ which is closer to $q_{\text{true}}$ for all but the lowest $N$ when compared to our formulations with $J'$.\footnote{Note that we have not optimised our $\alpha$ for minimising $J$ so it is not unreasonable to assume that the prior is dominating the solution for low $N$.}
The point evaluation approach therefore gives us \emph{consistent} point data assimilation when compared to the particular field reconstruction approach we test: it results in posterior consistency and is therefore consistent with Bayes' Theorem.

It is possible, were an l-curve analysis repeated for each $N$, that errors in our field reconstruction approach could be reduced. 
The lack of convergence would not change due to there being no mechanism for growing the misfit term with number of measurements.

\begin{figure}[hbtp]
    \centering
    \includegraphics[width=0.6\linewidth]{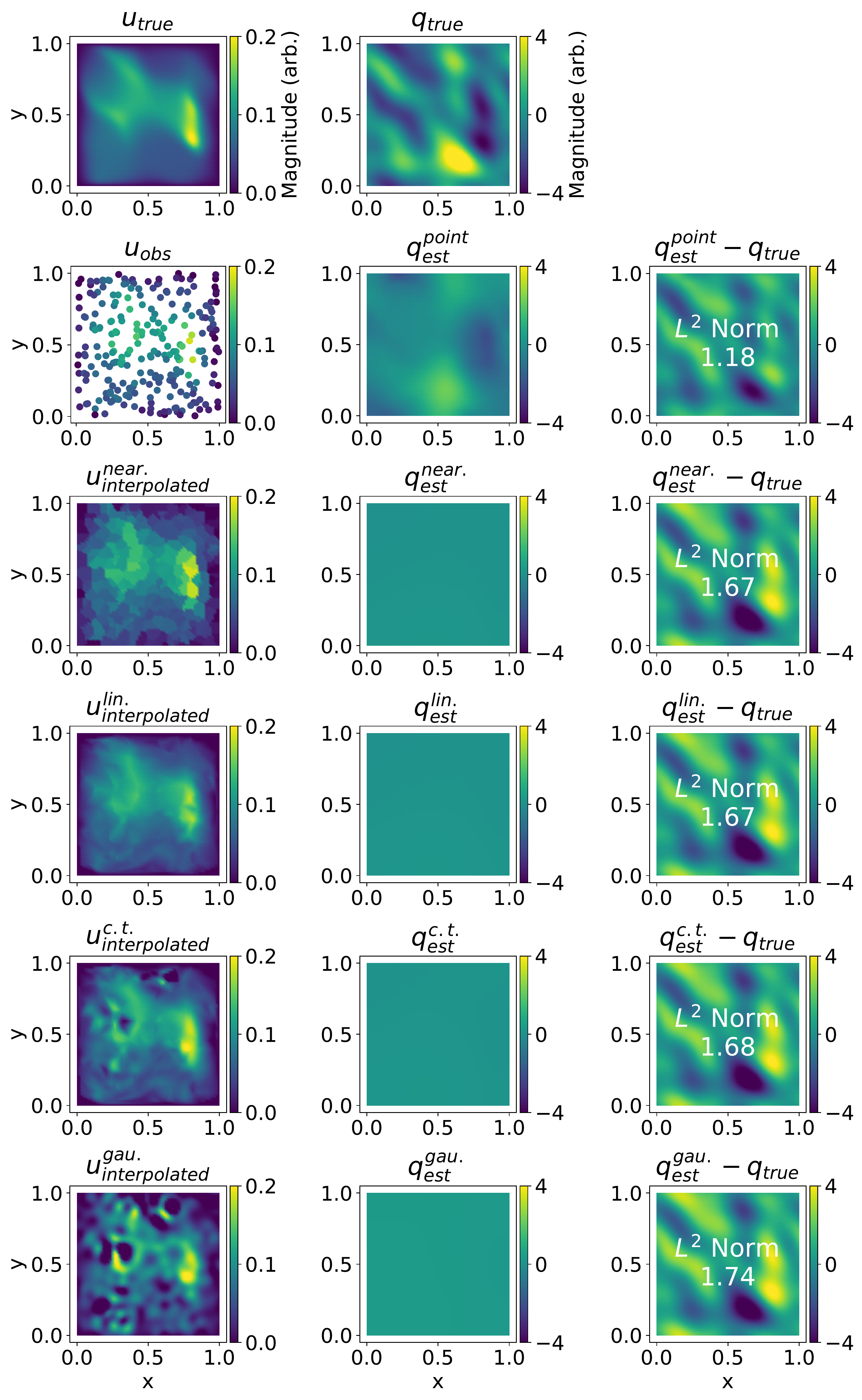}
    \caption{
        Summary plot of fields for $N = 256$.
        Rows correspond to method used where column 1 is the necessary $u$, column 2 is the corresponding $q$ at the optimum solution, and column 3 is the error.
        Row 1 shows the true $u$ and $q$.
        Row 2 has us minimising $J$ (Eq. \ref{eq:conductivity:J}) whilst rows 3-6 have us minimising $J'$ (Eq. \ref{eq:conductivity:Jprime}).
        The regularisation parameter $\alpha = 0.02$ throughout.
        The field we get after minimising $J$, $q_\text{est}^\text{point}$, manages to reproduce some features of $q_\text{true}$.
        For minimising $J'$ the solutions fail to reproduce any features of $q_\text{true}$ and the error is therefore higher.
        Each of the $u_{\text{interpolated}}$ fields are also visibly different from one another.
        For comparison see Fig. \ref{fig:conductivity:posterior:fields_32768}.
    }
    \label{fig:conductivity:posterior:fields_256}
\end{figure}

\begin{figure}[hbtp]
    \centering
    \includegraphics[width=0.6\linewidth]{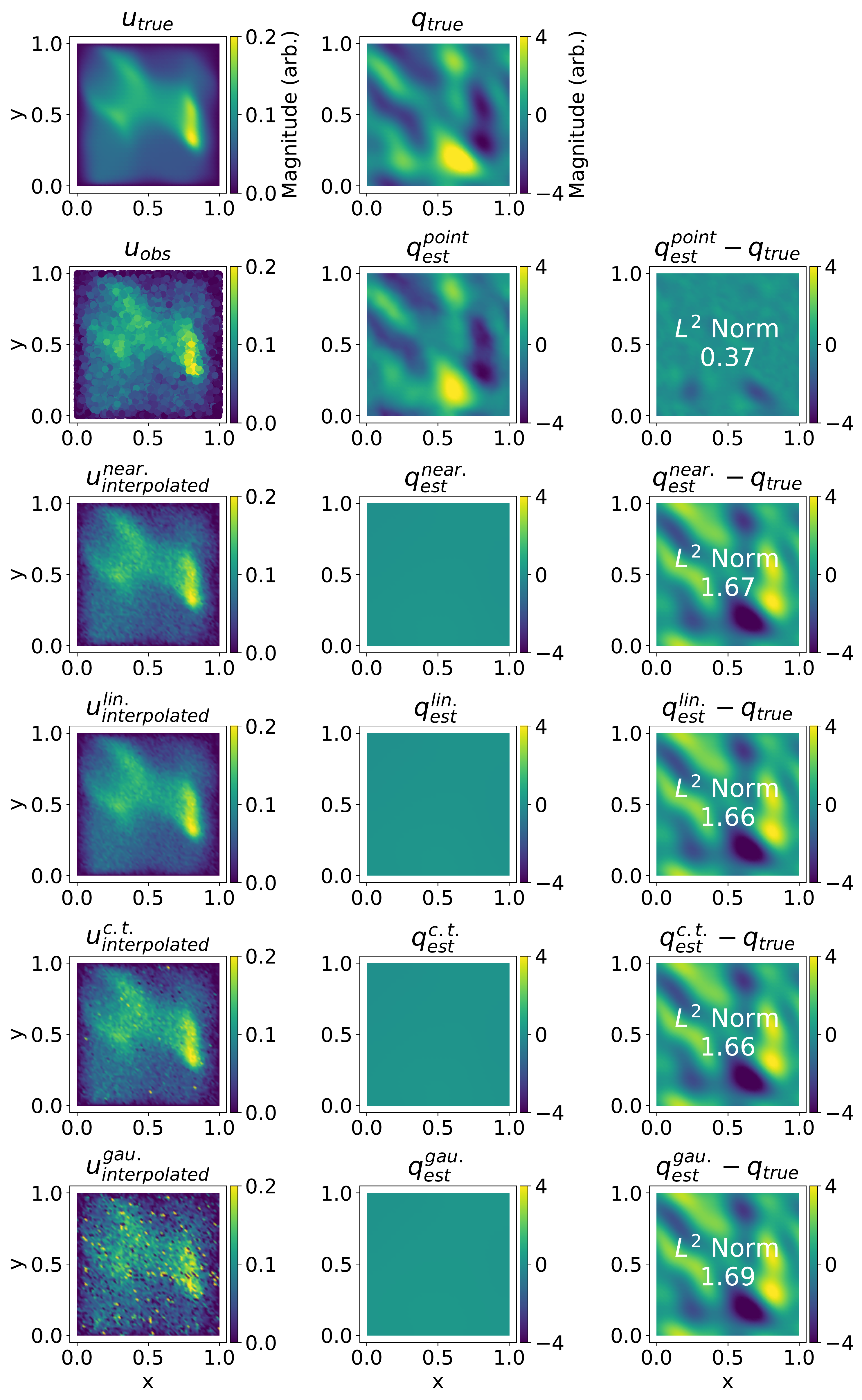}
    \caption{
        Summary plot of fields for $N = 32768$.
        Rows and columns correspond to those in Fig. \ref{fig:conductivity:posterior:fields_256}.
        The regularisation parameter $\alpha = 0.02$ throughout.
        We would expect the larger number of measurements to correspondingly reduce the error: this only occurs to a significant degree when solving $J$.
        This cannot be entirely blamed on a lack of mechanism in $J'$ for having the misfit term outgrow the regularisation term: the $u_{\text{interpolated}}$ fields do not approximate $u_\text{true}$ with $u_{\text{interpolated}}^{\text{gau.}}$ being particularly poor.
    }
    \label{fig:conductivity:posterior:fields_32768}
\end{figure}

We could attempt to enforce posterior consistency on the field reconstruction approach (minimising $J'$) by introducing a term in the model-data misfit which increases with the number of measurements.
Example calculated fields are shown in Fig. \ref{fig:conductivity:posterior:fields_256} and Fig. \ref{fig:conductivity:posterior:fields_32768}.
These demonstrate that the choice of interpolation method changes our field reconstruction. 
When attempting to enforce posterior consistency we would also need to ensure that our field reconstruction method approaches the true field as more measurement are performed.
There is no obvious way to do this which is universally applicable, particularly since measurements are always subject to noise.

\subsection{Groundwater hydrology} \label{sec:hydrology}

Our next example comes from groundwater hydrology.
The key field of interest is the \emph{hydraulic head} $\phi$, which has units of length.
In the following, we will apply the Dupuit approximation, which assumes that the main variations are in the horizontal dimension.

The first main equation of groundwater hydrology is the mass conservation equation
\begin{equation}
    S\frac{\partial\phi}{\partial t} + \nabla\cdot\mathbf u = q
\end{equation}
where $S$ is the dimensionless \emph{storativity}, $\mathbf u$ is the water velocity, and $q$ are the sources of water.
The second main equation is \emph{Darcy's law}, which states that the water velocity is proportional to the negative gradient of hydraulic head:
\begin{equation}
    \mathbf u = -T\nabla\phi
\end{equation}
where $T$ has units of length${}^2$ per unit time and is known as the \emph{transmissivity}.
The transmissivity is the product of the aquifer thickness and the hydraulic conductivity, the latter of which measures the ease with which water can percolate through the medium.
Clays have very low conductivity, sand and gravel much higher, and silty soil in between.

A typical inverse problem in groundwater hydrology is to determine the storativity and the transmissivity from measurements.
The measurements are drawn at isolated \emph{observation wells} where the hydraulic head can be measured directly.
To create a response out of steady state, water is removed at a set of discrete \emph{pumping wells}.

\begin{figure}
    \centering
    \includegraphics[width=0.49\linewidth]{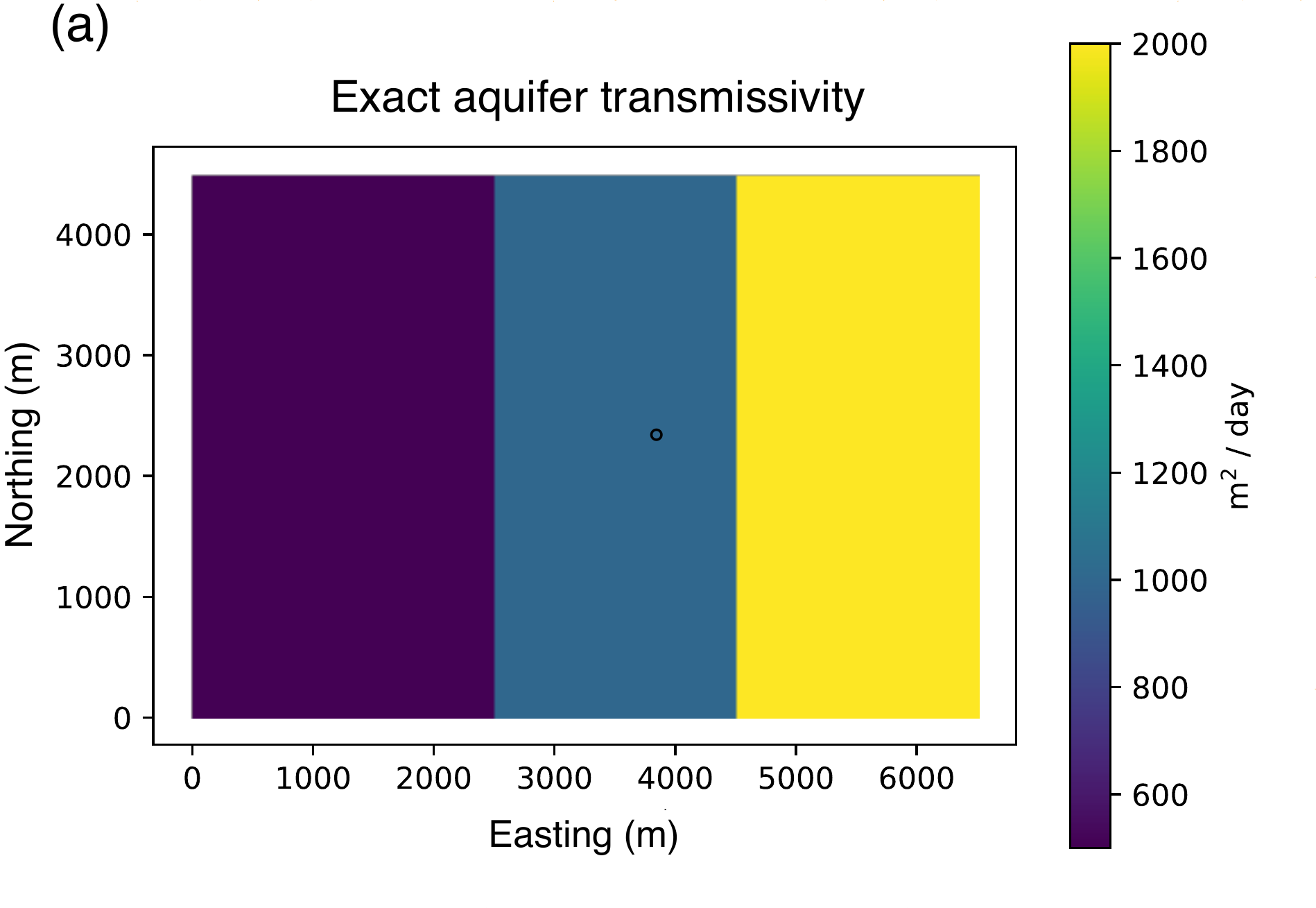}
    \includegraphics[width=0.49\linewidth]{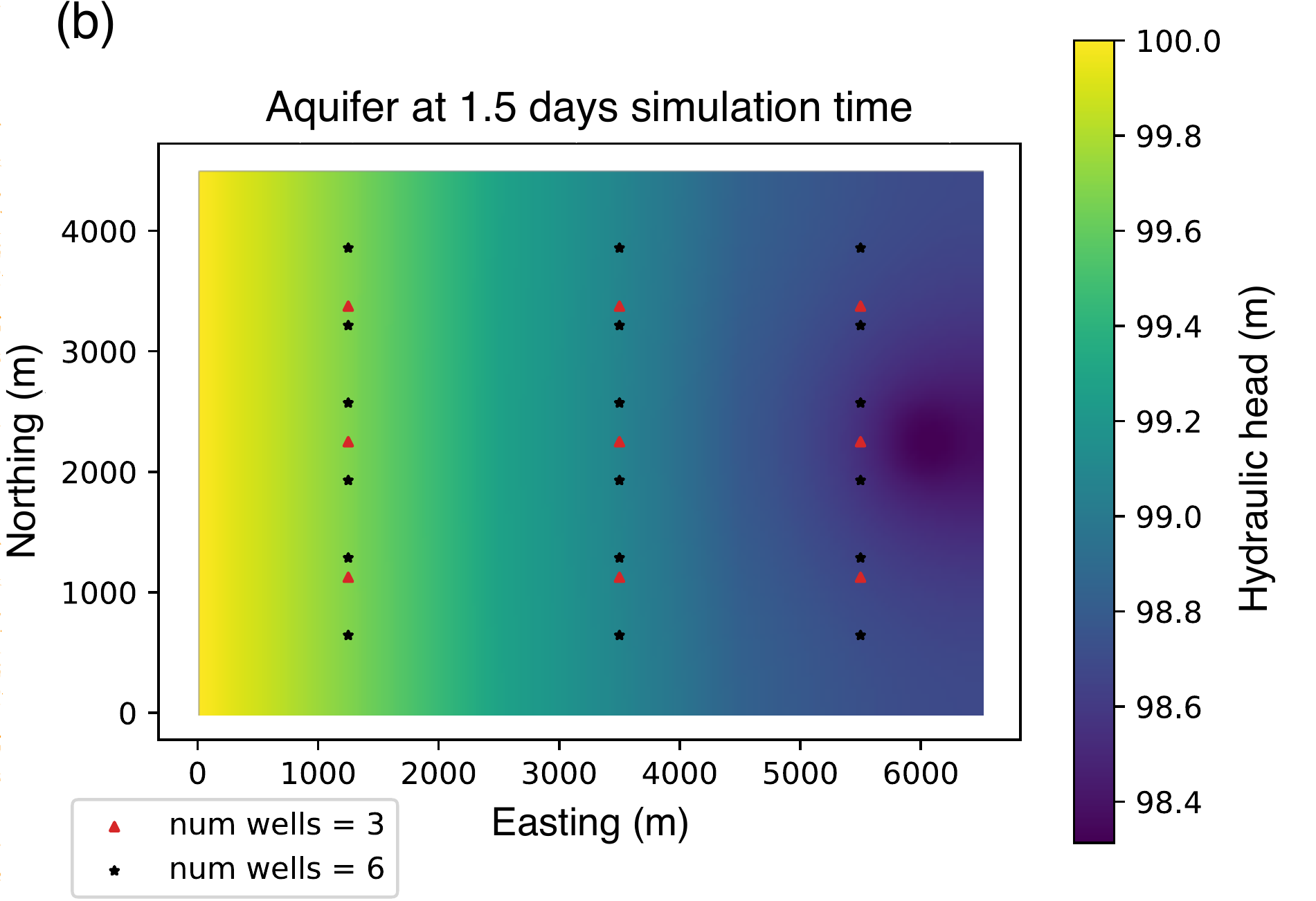}
    \caption{Exact transmissivity of the synthetic aquifer (a) and the final value of the hydraulic head (b).}
    \label{fig:hydrology-example-exact-solution}
\end{figure}

In the following, we will show a test case based on exercise 4.2.1-4.2.3 from  \citet{sun2013inverse}.
The setup for the model is a rectangular domain with the hydraulic head held at a constant value on the left-hand side of the domain and no outflow on the remaining sides.
The transmissivity is a constant in three distinct zones.
The hydraulic head is initially a uniform 100m and a pumping well draws 2000 m${}^3$ / day towards the right-hand side of the domain.
The exact value of the transmissivity and the final value of the hydraulic head are shown in Fig. \ref{fig:hydrology-example-exact-solution}.

\begin{figure}
    \centering
    \includegraphics[width=0.8\linewidth]{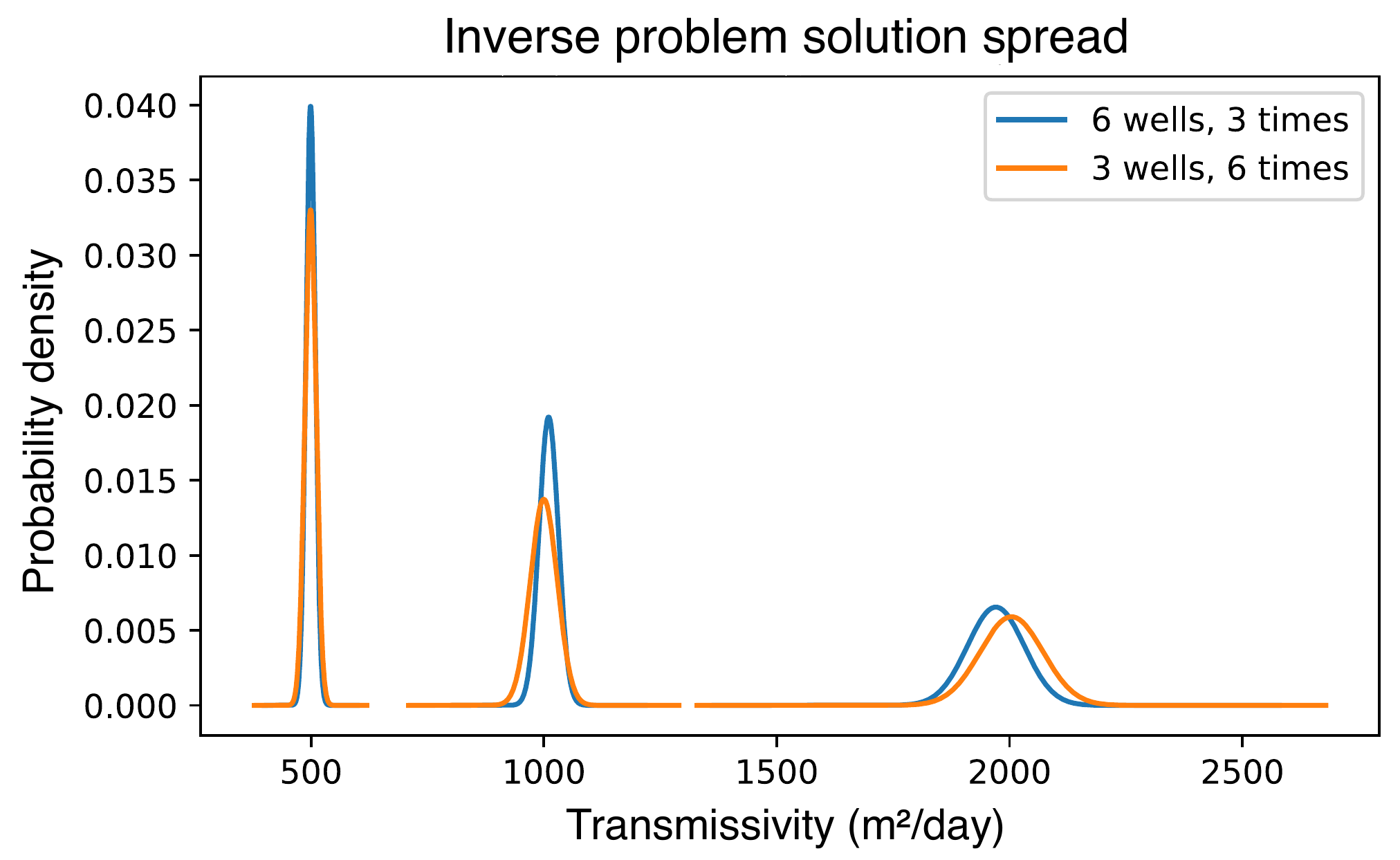}    \caption{Probability densities for the inferred transmissivities in each zone.
    The blue curves show the results obtained with a larger number of observation wells but fewer measurement times, while the orange curves show fewer observation wells but more measurement times.}
    \label{fig:transmissivity-probability-density}
\end{figure}

For the inverse problem, we examine two scenarios: (1) there are 6 observation wells in each zone that take measurements every 12 hours and (2) there are 2 observation wells in each zone that take measurements every 3 hours.
While this is a highly idealised problem, these kinds of experiments can inform real practice -- in this case, how to balance spatial and temporal density of measurements under limited resources.
The synthetic observations of the hydraulic head are centred at the true value with zero-mean normally distributed errors with a standard deviation of 1 cm.
We ran an ensemble of 30 simulations for each case.
Assuming the results are normally distributed, the probability densities for both scenarios are shown in Fig. \ref{fig:transmissivity-probability-density}.
In this case, we can observe that using more observation wells but fewer measurement times resulted in a smaller variance in the inferred transmissivity values than using the same number of total observations but with fewer observation wells and more measurement times.

The experiment above can only be conducted when the finite element modelling API includes support for assimilating point data.
In this case, the measurements are so sparse that they cannot be meaningfully interpolated to a densely-defined field.
Nonetheless, the unknown parameters are still identifiable.

\subsection{Ice shelves} \label{sec:ice_shelves}

\subsubsection{Physics model}

Our final example comes from glaciology; the main field of interest is the ice velocity.
On length and time scales greater than 100m and several days, glaciers flow like a viscous fluid.
The most principled equation set determining the velocity of a viscous fluid are the full Navier-Stokes equations, but the equation set we will work with uses several simplifications.
First, ice flow occurs at very low Reynolds number, and the ratio of the thickness to the length of the spatial domain is usually on the order of 1/20 or less.
Second, we will focus on \emph{ice shelves} -- areas where a glacier floats on the open ocean.
Most of the drainage basins of the Antarctic Ice Sheet terminate in floating ice shelves.
As far as the dynamics are concerned, ice shelves experience almost no friction at their beds.
As a consequence, the horizontal velocity is nearly constant with depth, so we can depth-average the equations.
The resulting PDE is called the \emph{shallow shelf equations}, which we describe below.
For complete derivations of all the common models used in glacier dynamics, see \citet{greve2009dynamics}.

The main unknown variable in the shallow shelf equations is the depth-averaged ice velocity $\mathbf u$, which is a 2D vector field.
The other key unknown is the ice thickness $h$.
Since an ice shelf is floating on the ocean, by matching the pressures at the base of the ice we find that the surface elevation of an ice shelf is $s = (1 - \rho_I / \rho_W)h$ where $\rho_I$, $\rho_W$ are respectively the density of ice and ocean water.

A key intermediate variable is the \emph{membrane stress tensor}, which we will write as $\mathbf M$.
The membrane stress tensor has units of stress and has rank 2, i.e. it is a 2 $\times$ 2 matrix field.
Physically, the membrane stress plays the same role for this simplified 2D problem as the full stress tensor does for the full Stokes equations.
The shallow shelf equations are a conservation law for membrane stress:
\begin{equation}
    \nabla \cdot (h\mathbf M) - \frac{1}{2}\rho_I(1 - \rho_I / \rho_W)g\nabla h^2 = 0
    \label{eq:ice:membrane-stress-conservation}
\end{equation}
where $g$ is the acceleration due to gravity.

To obtain a closed system of equations, we need to supply a \emph{constitutive relation} -- an equation relating the membrane stress tensor to the depth-averaged velocity.
First, the \emph{strain rate tensor} is defined to be the rank-2 tensor
\begin{equation}
    \boldsymbol{\dot\varepsilon} = \frac{1}{2}\left(\nabla \mathbf u + \nabla\mathbf u^\top\right),
    \label{eq:ice:strain-rate}
\end{equation}
i.e. the symmetrized gradient of the depth-averaged velocity.
The membrane stress tensor is then proportional to the strain rate tensor:
\begin{equation}
    \mathbf M = 2\mu\left(\boldsymbol{\dot\varepsilon} + \text{tr}(\boldsymbol{\dot\varepsilon})\mathbf I\right)
    \label{eq:ice:membrane-stress-definition}
\end{equation}
where $\mathbf I$ is the 2 $\times$ 2 identity tensor and $\mu$ is the viscosity coefficient, which has units of stress $\times$ time.

One of the more challenging parts about glacier dynamics is that the viscosity also depends on the strain rate.
This makes the shallow shelf equations nonlinear in the velocity.
The most common assumption is that the viscosity is a power-law function of the strain rate tensor:
\begin{equation}
    \mu = \frac{A^{-\frac{1}{n}}}{2}\sqrt{\frac{\boldsymbol{\dot\varepsilon} : \boldsymbol{\dot\varepsilon} + \text{tr}(\boldsymbol{\dot\varepsilon})^2}{2}}^{\frac{1}{n} - 1}
    \label{eq:ice:viscosity-definition}
\end{equation}
where $n$ is an exponent and $A$ is a prefactor called the \emph{fluidity}.
Laboratory experiments and field observations show that $n \approx 3$; this is referred to as \emph{Glen's flow law} \citep{greve2009dynamics}.
The fact that $n > 1$ makes ice a \emph{shear-thinning} fluid, i.e. the resistance to flow decreases at higher strain rate.
The fluidity coefficient $A$ has units of stress${}^n$ $\times$ time.
(This unit choice and the exponent of $-1/n$ on $A$ in equation \eqref{eq:ice:viscosity-definition} reflects the fact that historically glaciologists have, by convention, written the constitutive relation as an equation defining the strain rate as a function of the stress.
For solving the momentum balance equations, we have to invert this relation.)
Several factors determine the fluidity, the most important of which is temperature -- warmer ice is easier to deform.

Putting together equations \eqref{eq:ice:membrane-stress-conservation}, \eqref{eq:ice:strain-rate}, \eqref{eq:ice:membrane-stress-definition}, \eqref{eq:ice:viscosity-definition}, we get a nonlinear second-order elliptic PDE for $\mathbf u$.
The last thing we need to complete our description of the problem is a set of boundary conditions.
We fix the ice velocity along the inflow boundary, i.e. a Dirichlet condition.
Along the outflow boundary, we fix the normal component of the membrane stress:
\begin{equation}
    h\mathbf M\cdot\boldsymbol\nu = \frac{1}{2}\rho_I(1 - \rho_I / \rho_W)gh^2\boldsymbol\nu
\end{equation}
where $\boldsymbol\nu$ is the unit outward-pointing normal vector to the boundary of the domain.
This is a Neumann-type boundary condition.

\subsubsection{Inverse problem}

The key unknowns in the shallow shelf equations are the thickness, velocity, and fluidity.
Both the ice velocity and thickness are observable at large scales using satellite remote sensing.
The fluidity is not directly measurable at large scales.
The goal of our inverse problem is to estimate the fluidity from measurements of ice velocity.

As a test case, we will look at the Larsen C Ice Shelf in the Antarctic Peninsula.
We will use the BedMachine map of Antarctic ice thickness \citep{morlighem2020deep} and the MEaSUREs InSAR phase-based velocity map \citep{mouginot2019continent}.

To ensure positivity of the fluidity field that we estimate, we will, as in Sect. \ref{sec:unknown_conductivity}, write
\begin{equation}
    A = A_0e^\theta
\end{equation}
and infer the dimensionless log-fluidity field $\theta$ instead.

The first paper to consider inverse problems or data assimilation in glaciology was \citet{macayeal_basal_1992} which referred to the formulation of a functional to be minimised as the \emph{control method}.
Since then, most of the work in the glaciology literature on data assimilation has assumed that the observational data can be interpolated to a continuously-defined field and used as a misfit functional the 2-norm difference between the interpolated velocity and the computed velocity \citep{joughin2004basal, vieli_numerical_2006, shapero2016basal}.
Assuming that the target velocity field to match is defined continuously throughout the entire domain, however, obscures the fact that there are only a finite number of data points.
A handful of publications have taken the number of observations into account explicitly in order to apply further statistical tests on goodness-of-fit, for example \citet{macayeal1995basal}.
We argue that making it possible to easily assimilate sparse data will improve the statistical interpretability of the results.
Moreover, one of the main uses for data assimilation is to provide an estimate of the initial state of the ice sheet for use in projections of ice flow and extent into the future.
Improving the statistical interpretability of the results of glaciological data assimilation will help to quantify the spread in model projections due to uncertainty in the estimated initial state.

To demonstrate the capabilities of the new point data assimilation features in Firedrake, we will conduct a type of \emph{cross-validation} experiment.
The idea of cross-validation is to use only a subset of the observational data $u^o$ to estimate the unknown field.
The data that are held out are then used to determine the goodness of fit.
Rather than use, for example, the Morozov discrepancy principle \citep{habermann2012reconstruction} or the L-curve method to select the regularisation parameter, we can choose it as whichever value gives the smallest misfit on the held-out data.
This experiment would be meaningless using the field reconstruction approach because, at points where we have held out data, we would be matching the computed velocity field to interpolated values, when the whole point of the exercise is to not match the computed velocity field to anything where we have no data.
A common approach is to leave out only a single data point; for linear problems, there are analytical results that allow for much easier estimation of the best value of the regularisation parameter using leave-one-out cross-validation \citep{picard1984cross}.
Here we will instead randomly leave out some percentage of the observational data.

Formally, let $\{u^o(x_k)\}$ be a set of $N$ observations of the velocity field at points $\{x_k\}$.
Let $f$ be some parameter between 0 and 1 and select uniformly at random a subset $I$ of size $f \times N$ of indices between 1 and N.
The model-data misfit functional for our problem is
\begin{equation}
    E(u) = \sum_{k \in I}\frac{|u(x_k) - u^o(x_k)|^2}{2\sigma_k^2}
\end{equation}
where $\sigma_k$ is the formal estimate of the error of the $k$th measurement as reported from the remote sensing data.
Note that we sum over only the data points in $I$ and not all of the observational data.
The regularisation functional is, again similarly to Sect. \ref{sec:unknown_conductivity},
\begin{equation}
    R(\theta) = \frac{\alpha^2}{2}\int_\Omega|\nabla\theta|^2 dx.
\end{equation}
We can then minimise the functional 
\begin{equation}  \label{eq:ice:minimization_problem}
    E(u) + R(\theta)   
\end{equation}
subject to the constraint that $u$ is a solution of the momentum balance equations \ref{eq:ice:membrane-stress-conservation}, \ref{eq:ice:strain-rate}, \ref{eq:ice:membrane-stress-definition}, \ref{eq:ice:viscosity-definition} for the given $\theta$.

We now let $\theta_\alpha$, $u_\alpha$ be the log-fluidity and velocity obtained by minimising Eq. \ref{eq:ice:minimization_problem} using a regularisation parameter $\alpha$.
We will then compute the misfit against the held-out data:
\begin{equation}
    E'(\alpha) = \sum_{k \notin I}\frac{|u_\alpha(x_k) - u^o(x_k)|^2}{2\sigma_k^2}.
\end{equation}
The key difference here is that we sum over the indices \emph{not} in the training set ($k \notin I$) instead of those in the training set ($k \in I$).
The right choice of $\alpha$ is the minimiser of $E'$.

\begin{figure}
    \centering
    \includegraphics[width=0.49\linewidth]{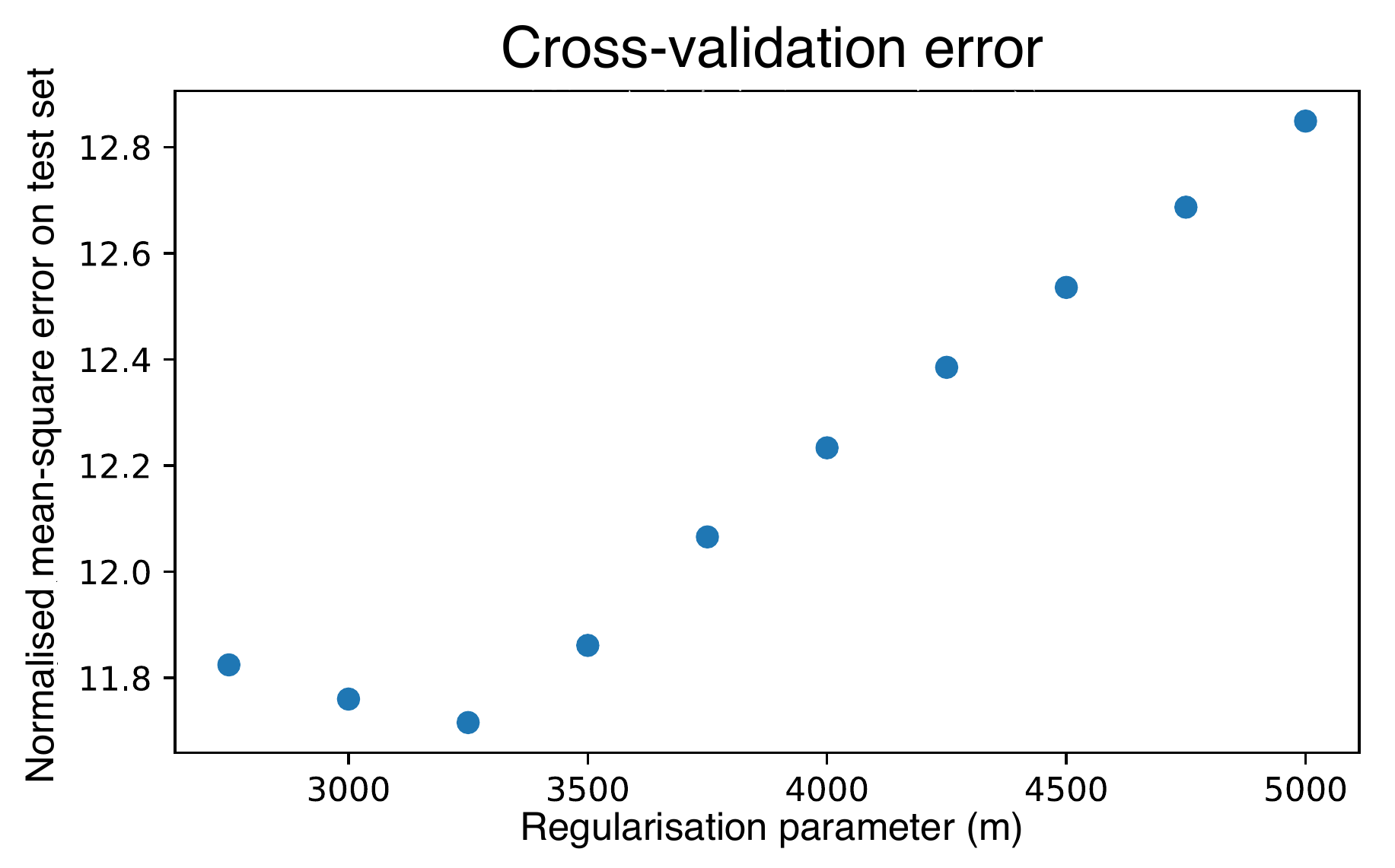} \includegraphics[width=0.49\linewidth]{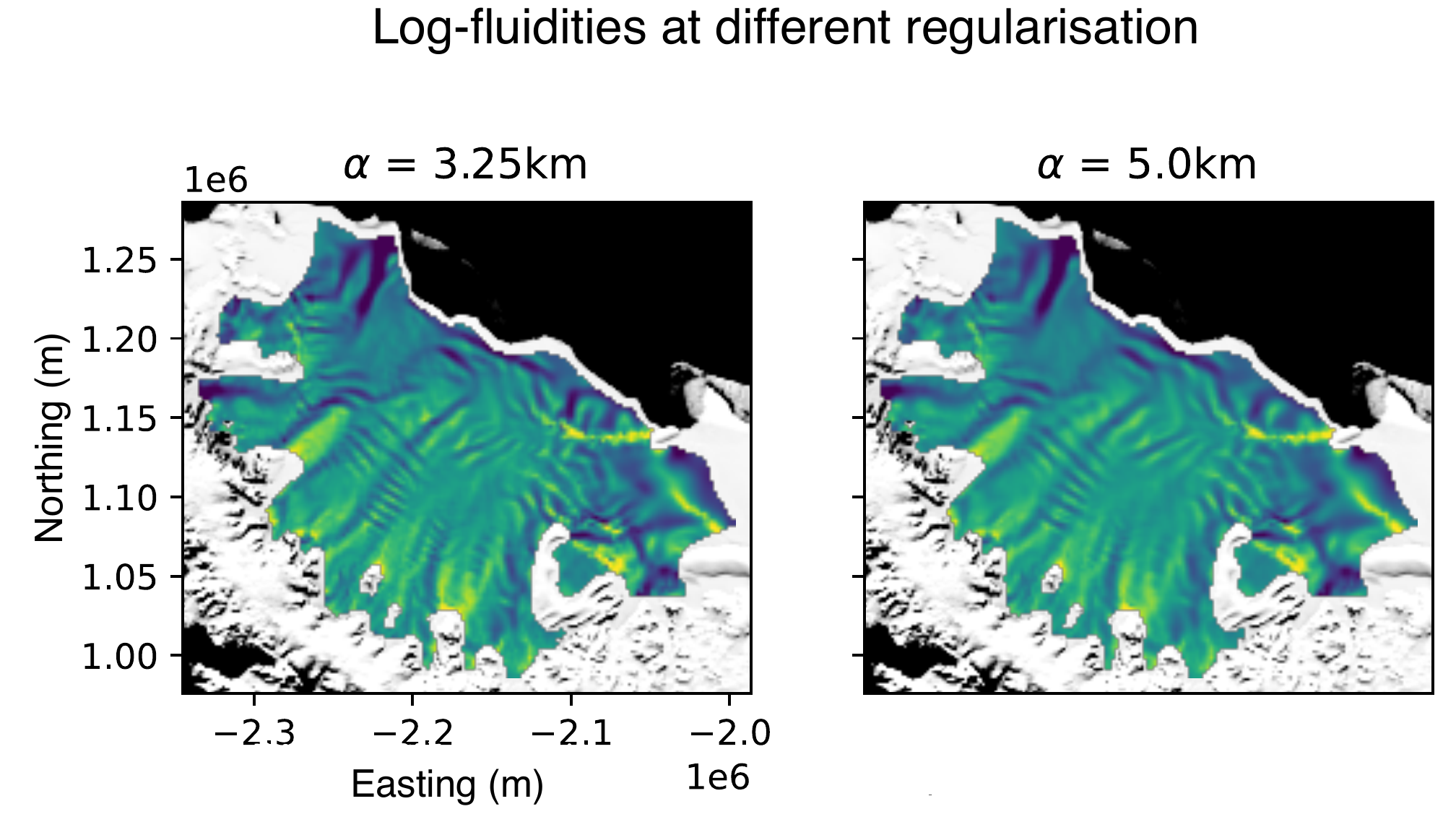}
    \caption{Normalized cross-validation errors ($E'(\alpha)$) and inferred log-fluidity fields with regularisation parameter set to 3.25 km and 5km.
    The log-fluidity background image is the MODIS Mosaic of Antarctica \citep{scambos_modis-based_2007, haran_modis_2021}, courtesy of the NASA National Snow and Ice Data Center (NSIDC) Distributed Active Archive Center (DAAC).}
    \label{fig:xval}
\end{figure}

This experiment was performed using Icepack \citep{shapero2021icepack}, which is built on top of Firedrake.
As our target site, we used the Larsen C Ice Shelf in the Antarctic Peninsula.
There are a total of 235,510 grid points within the domain; we used 5\% or 11,775 of these as our training points.
The results of the cross-validation experiment are shown in Fig. \ref{fig:xval}.
We found a well-defined minimum at $\alpha$ = 3.25 km.
Fig. \ref{fig:xval} shows the log-fluidity fields obtained at 3.25 km, which is the appropriate regularisation level, and 5 km, which is over-regularised.
With $\alpha$ = 5 km, several features are obscured or blurred out.

The remote sensing products that we used here come with formal estimates for the standard errors $\sigma_k$, but these formal error estimates might not be truly indicative of the actual errors.
Cross-validation allows us to estimate the measurement errors after the fact if we assume that there is some uniform scaling constant $c$ such that $\sigma_k^{\text{true}} = c \times \sigma_k$.
In our experiment, the normalised sum of squared cross-validation errors at the appropriate value of the regularisation parameter ($\alpha = 3$km) was roughly 11.8.
If our model physics are accurate this suggests that the formal errors under-estimate the true errors by a factor of about 3.4 by taking square roots.
Pinning down the degree to which the formal errors under- or over-estimate the true errors is difficult by any other means.
The Morozov discrepancy principle, on the other hand, assumes that the measurement errors used in the objective functional are the true measurement errors \citep{habermann2012reconstruction}.
The discrepancy principle breaks down when its assumptions are violated.

The goal of this experiment is to demonstrate the capability of the point data assimilation features in Firedrake to do something that was not possible with existing modelling tools.
We do not assert that all future work in glaciological data assimilation should necessarily use a cross-validation type approach for regularisation parameter selection instead of the L-curve, the discrepancy principle, or other related methods.
It is possible that cross-validation, the discrepancy principle, and the L-curve all happen to give roughly similar results on most problems of interest, in which case the best choice is the cheapest computationally.
But a comparison is only possible with proper point data assimilation.

A sufficiently good optimisation algorithm can find a minimiser of any reasonable functional that a modeller would care to throw at it.
Without checks on whether the results are reasonable or not, such as tests from statistical estimation theory on how good a fit one can expect given the errors in the remote sensing data, there is no way to discern whether the computed minimiser is meaningful or a mere artefact.
Arguably, the more interesting case for glaciologists is when the inferred field does \emph{not} fit the observational data.
Assuming that any regularisation employed had a small effect then this points to either (1) a failure of the physics model that must be corrected by improving our understanding of glacier dynamics or (2) a failure of the remote sensing methodology that must be corrected by improving our understanding of how the raw instrument data gets turned into higher-level, physically meaningful information.

\section{Future Work}

The next step is to implement moving points.
The underlying concepts of interpolating from a parent mesh onto a vertex-only mesh remain unchanged: the vertex-only mesh would now move over time.

Moving points could be used for assimilating data from Lagrangian points, such as data from weather buoys which follow ocean currents.
For static experiments, we could include uncertainty in the location of measurements through a differentiable point relocation operator.
Moving points could allow particle in cell methods to be introduce to Firedrake as has been done for FEniCS \citep{maljaars_leopart_2021}.

Full cross-mesh parallel compatible interpolation onto static and moving meshes in Firedrake ought to be possible by extending this work.\footnote{Prior work on supermeshing \citep{maddison_directional_2012, farrell_conservative_2009} provides a mathematical framework.}
This would significantly improve the input and output capabilities of Firedrake.
For data input, we could perform data assimilation of arbitrary field data: this could come from measurements of be the output of another model.
For output we could interpolate fields into domains which match another model or data set.
We could use this to calculate fluxes and volume integrals in much the same way as we are now able to calculate point evaluations.
Terms in PDEs which are defined at particular locations in a domain (i.e. similar to Eq. \ref{eq:interpolation:delta_equiv} but over extended domains) could also be directly represented in Firedrake without having to be aligned to our mesh.

\conclusions  

This work makes two key points.
Firstly, a finite element representation of point data enables the automation of point evaluations.
This, in turn, lets us automatically solve PDE-constrained optimisation problems where our functionals contain point evaluations.

Secondly, the use of point evaluation misfit functionals instead of field reconstruction approaches in data assimilation has several benefits: it (i) ensures our inverse problem displays posterior consistency, (ii) reduces errors, (iii) avoids the need for ambiguously defined inter-measurement interpolation regimes, (iv) allows assimilation of very sparse measurements and (v) enables more statistical analysis.

The second result is both a demonstration of new functionality and a general call for all scientific communities who face these kinds of inverse problems to carefully consider if point evaluation misfit functionals would be appropriate for their use case.
This is particularly salient for users of finite element methods where point evaluation of fields is always well defined over the whole domain.
For data assimilation problems where this is not possible, ensuring that the model-data misfit grows with the number of measurements could also be considered.

\appendix
\section{L curves}


\begin{figure}[hbtp]
    \centering
    \includegraphics[width=\linewidth]{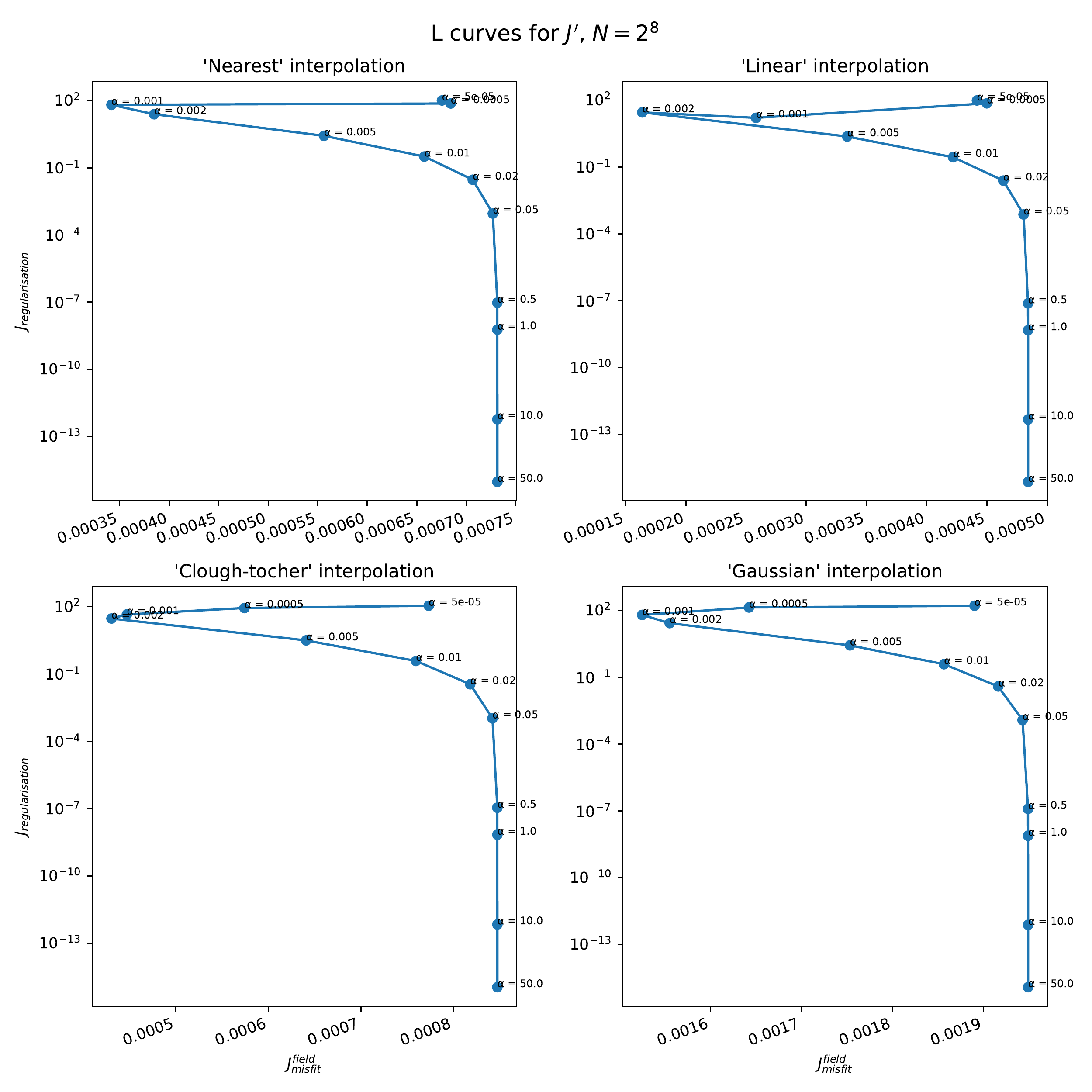}
    \caption{
        L-curves from minimising $J'$ (Eq. \ref{eq:conductivity:Jprime}) with different methods for estimating $u_{\text{interpolated}}$.
        For low $\alpha$, $J^{\text{field}}_{\text{misfit}}$ stopped being minimised and solver divergences were seen for $u_{\text{interpolated}}^{\text{gaussian}}$.
        The problem was likely becoming overly ill formed and the characteristic `L' shape (with sharply rising $J_{\text{regularisation}}$ for low $\alpha$ and a tail-off for large $\alpha$) is therefore not seen.
        To keep the problem well formed without the regularisation parameter being too big $\alpha = 0.02$ is chosen for each method.
        }
    \label{fig:conductivity:l-curves-Jprime}
\end{figure}
\clearpage

\begin{figure}[hbtp]
    \centering
    \includegraphics[width=0.49\linewidth]{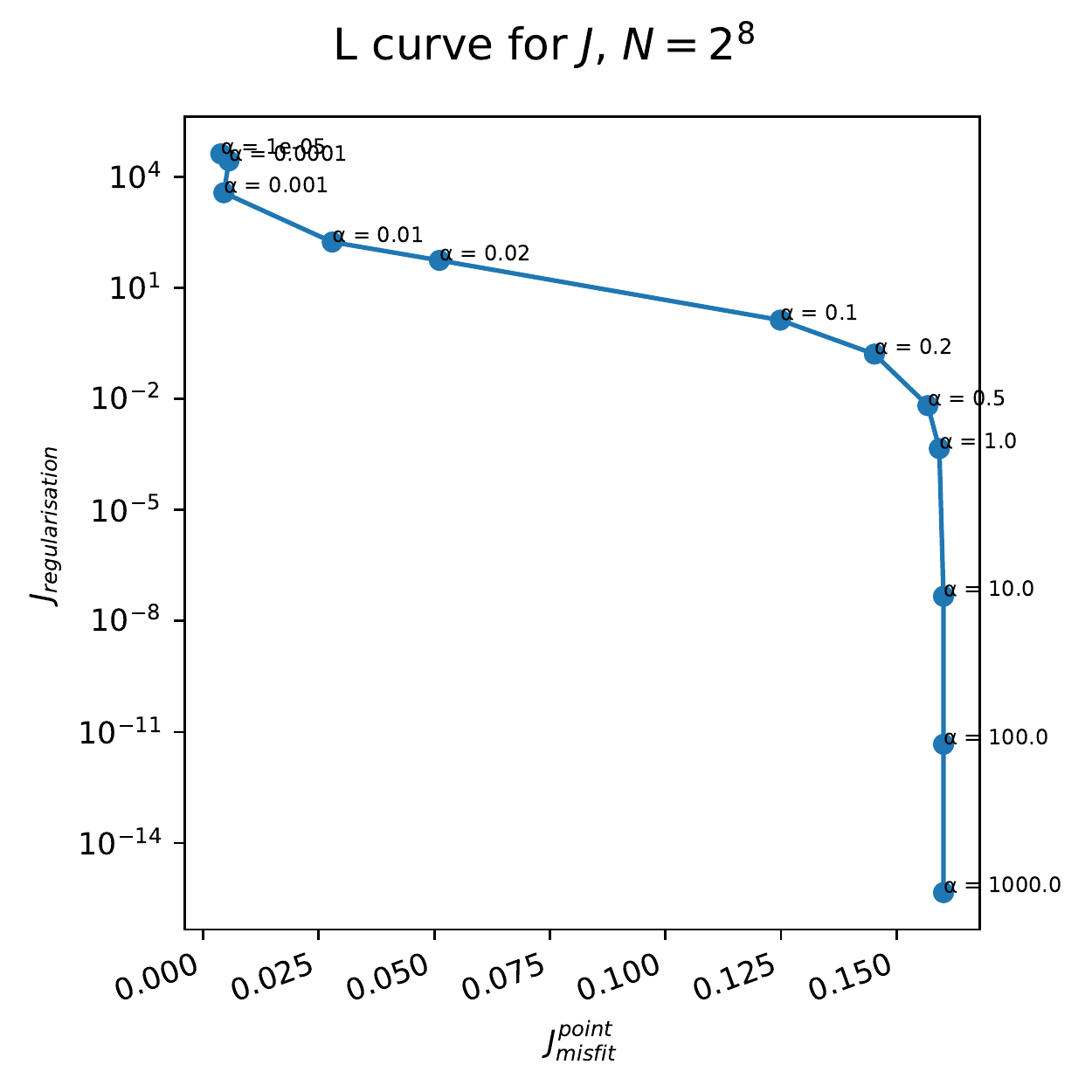}
    \caption{
        L-curve for minimising $J$ (Eq. \ref{eq:conductivity:J}).
        The characteristic `L' shape is seen and $\alpha = 0.02$ is seen to be close to the turning point and is therefore chosen for consistency with the other L-curves (Fig. \ref{fig:conductivity:l-curves-Jprime}).
        }
    \label{fig:conductivity:l-curves-J}
\end{figure}
\clearpage

\noappendix

\codedataavailability{
The demonstrations use publicly available code in a GitHub repository which is archived with Zenodo \citep{reuben_w_nixon_hill_2023_7950441}.
All figures can be generated using that repository.
The version of Firedrake used for the unknown conductivity demonstration is archived on Zenodo \citep{zenodo/Firedrake-20230316.0}.
The version of Firedrake used for the groundwater hydrology and ice shelf demonstration is similarly archived \citep{zenodo/Firedrake-20230405.1}.
The ice shelf demonstration uses Icepack \citep{shapero2021icepack}, specifically this Zenodo archived version \citep{daniel_shapero_2023_7897023}.

The BedMachine thickness map \citep{morlighem2020deep, morlighem2022measures} and the MEaSUREs InSAR phase-based velocity map \citep{mouginot2019continent} used in the ice shelf demo are publicly hosted at the US National Snow and Ice Data Center.
}

\authorcontribution{
RWNH and DAH formulated the point data point evaluation ideas with CJC validating the mathematics.
RWNH made software modifications to Firedrake \citep{rathgeber_firedrake_2016}, FInAT \citep{homolya_exposing_2017}, FIAT \citep{kirby_algorithm_2004}, UFL \citep{alnaes_unified_2014} and TSFC \citep{homolya_tsfc_2018} necessary for this work under the supervision of DAH.
DS and RWNH were responsible for the unknown conductivity experiment methodology and validation; RWNH performed the experiment and was responsible for visualisation.
DS was responsible for methodology of the groundwater hydrology and ice shelf experiments with RWNH performing validation; DS and RWNH were responsible for visualisations.
DS is the author of the Icepack \citep{shapero2021icepack} software package.
RWNH and DS wrote the initial draft presentation.
All authors contributed to review and editing.
}

\competinginterests{
David A. Ham, a coauthor of this work, is the chief-executive editor of the Geoscientific Model Development journal.
}

\begin{acknowledgements}
This work was supported by the Natural Environment Research Council [NE/S007415/1].
DRS is funded by the US National Science Foundation (grant \#1835321) and National Aeronautics and Space Administration (grant \#80NSSC20K0954).
\end{acknowledgements}

\bibliographystyle{copernicus}  
\bibliography{references, references_zotero_rnh}

\end{document}